\documentclass[aps,prd,linenumber,twocolumn,nomerge,superscriptaddress,preprintnumbers,superscriptaddress,nofootinbib,floatfix,showkeys]{revtex4-2}

\usepackage{graphicx}
\usepackage{xcolor}
\usepackage{orcidlink}
\usepackage{dcolumn}
\usepackage{hyperref}
\usepackage{amssymb}
\usepackage{xspace}
\usepackage{amsmath}
\usepackage{cleveref}
\usepackage{soul}
\usepackage[mathlines]{lineno}
\usepackage{setspace} 
\usepackage[mathlines]{lineno}

\newcommand{\Msol}{\xspace\rm{M}_{\odot}\xspace}

\newcommand{\icarogw}{\textsc{icarogw}\xspace} 
 
\newcommand{\de}{{\rm d}}

\newcommand{\Ngw}{\ensuremath{N_{\rm GW}}\xspace}

\newcommand{\Ncbc}{\ensuremath{N_{\rm CBC}}\xspace}
\newcommand{\Nexp}{\ensuremath{N_{\rm exp}}\xspace}
\newcommand{\Pdet}{\ensuremath{P_{\rm det}}\xspace}

\newcommand{\Tobs}{\ensuremath{T_{\rm obs}}\xspace}

\newcommand{\nn}{\nonumber}

\renewcommand{\vec}[1]{\ensuremath{\overrightarrow{#1}}}

\begin{document}

\preprint{Short-author PP}

\title{The Spin Magnitude of Stellar-Mass Binary Black Holes Evolves with the Mass: Evidence from Gravitational Wave Data}

\author{Grégoire Pierra \orcidlink{0000-0003-3970-7970}}
\email{g.pierra@ip2i.in2p3.fr}
\affiliation{Universite Claude Bernard Lyon 1, CNRS/IN2P3, IP2I Lyon, UMR 5822, Villeurbanne, F-69100, France}

\author{Simone Mastrogiovanni \orcidlink{0000-0003-1606-4183}}
\affiliation{INFN, Sezione di Roma, I-00185 Roma, Italy}

\author{Stéphane Perriès \orcidlink{0000-0003-2213-3579}}
\affiliation{Universite Claude Bernard Lyon 1, CNRS/IN2P3, IP2I Lyon, UMR 5822, Villeurbanne, F-69100, France}

\date{\today}

\begin{abstract}
The relation between the mass and spin of stellar-mass binary black holes (BBHs) has been proposed to be a smoking gun for the presence of multiple formation channels for compact objects. First-generation black holes (BHs) formed by isolated binary stellar progenitors are expected to have nearly aligned small spins, while nth-generation BBHs resulting from hierarchical mergers are expected to have misaligned and higher spins.
Leveraging data from the third observing run O3 (GWTC-2.1 and GWTC-3), we employ hierarchical Bayesian methods to conduct a comprehensive study of possible correlations between the BBH masses and spins. We use parametric models that either superpose independent BBH populations or explicitly model a mass-spin correlation. We unveil strong evidence for a correlation between normalized spin magnitudes and masses of BBHs. The correlation can be explained as a transition from a BBH population with low spins at low masses and higher spins for higher masses. Although the spin magnitude distribution at high masses lacks robust constraints, we find strong evidence that a transition between two BBH populations with different spin distributions should happen at 40-50 $M_\odot$. In particular, we find that the population of BBHs above  40-50 $M_\odot$ should compose the $\sim 2\%$ of the overall population, with a spin magnitude $\chi$ peaking around 0.7, consistently with the fraction of nth-generation BBHs formed by hierarchical mergers in the latest state-of-the-art BBH genesis simulations.
\end{abstract}

\maketitle

\newpage

The mechanisms governing the formation of stellar-mass binary black holes (BBHs) are still under debate. BBHs are believed to originate from three main formation channels: isolated evolution of stellar binaries, dynamical assembly, or hierarchical mergers \cite{Mapelli:2020vfa, Mapelli:2021taw, MANDEL20221}. These channels significantly influence various properties of BBH binaries such as their masses, spins and eccentricity. For instance, the possible correlation between the BBH spin magnitudes and masses has been proposed as a possible smoking gun for the presence of diverse formation channels such as BBHs formed by isolated stellar binaries, dynamically assembled in dense stellar environments and from hierarchical mergers \cite{Wen:2002km, Marchant:2016wow, Bartos:2016dgn, Mapelli:2020vfa, Bouffanais:2021wcr, Mapelli:2021taw}.
As an illustration, $1^{st}$-generation BHs formed from isolated stellar binaries are expected to have masses $\lesssim 50 M_\odot$, the Pair Instability Supernova Gap (PISN) \cite{Farmer:2019jed}, and relatively small spins aligned to the orbital angular momentum. Instead, $n^{th}$-generation BHs born by previous mergers and in binaries formed in dense stellar environments, are expected to have higher spins\footnote{Possibly around 0.7, from the pre-merger orbital angular momentum \cite{Gerosa:2021mno}.} and misaligned with respect to the orbital angular momentum.

Since the first detection of gravitational waves (GWs) in 2015, originating from the merger of two stellar-mass black holes (BHs) \cite{LIGOScientific:2016aoc}, astrophysicists have gained a powerful probe to study these systems in greater detail. GWs have emerged as an invaluable means to directly examine and understand the astrophysical characteristics of BBH populations and the intricate influence of their environments on formation processes. The release of the largest catalog of GW events to date, GWTC-2.1 and GWTC-3, in 2021 by the  LIGO-Virgo-KAGRA Collaboration (LVK) has further advanced our understanding of BBH populations \cite{KAGRA:2013rdx, KAGRA:2021vkt}. The latest catalog includes 90 detected events, with the majority originating from BBH mergers. Subsequent population inference studies have studied the BBH population properties using parametric and non-parametric models utilizing Bayesian inference \cite{Vitale:2020aaz, Ashton:2018jfp, Thrane:2018qnx, KAGRA:2021duu, Rinaldi:2021bhm, Cheng:2023ddt, Farah:2024xub}. The latest population results from GWTC-3 find that \cite{KAGRA:2021duu}: \textit{(i)} the distribution of the spin magnitudes of BBHs prefers lower values of the spins $\chi \lesssim 0.4$ and there is no compelling evidence for a subpopulation of BBHs with zero spins \cite{Kimball:2020opk, Callister:2022qwb,Tong:2022iws, Mould:2022xeu,Adamcewicz:2023szp} \textit{(ii)} the BBH spin component aligned to the orbital angular momentum does not significantly correlate with the mass of the object, \textit{(iii)}  higher spins correlate with asymmetric mass binaries \cite{Callister:2021fpo, Adamcewicz:2022hce, Adamcewicz:2023mov}.

\begin{figure}
    \centering
    \includegraphics[width=.5\textwidth]{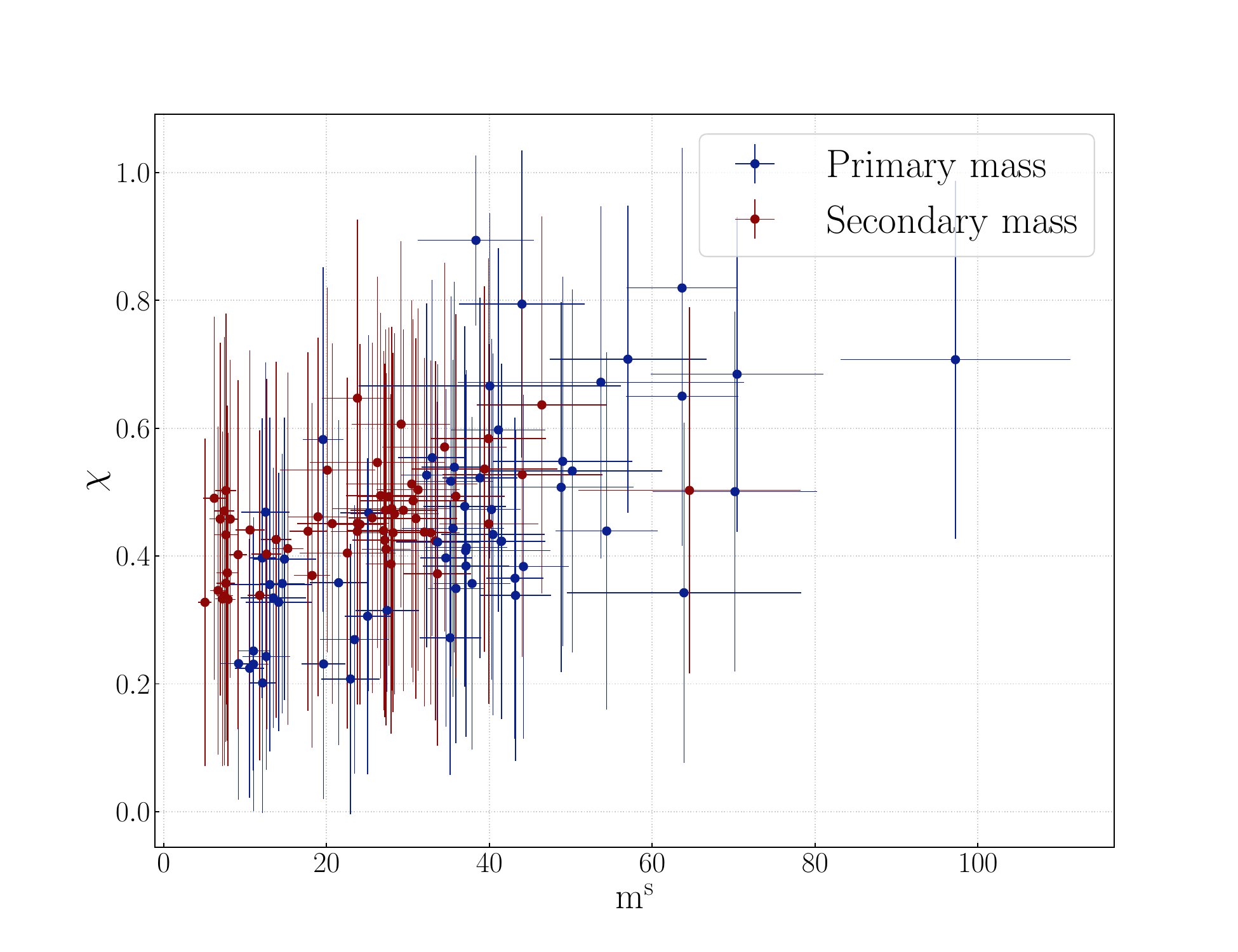}
    \caption{\textbf{Mass-Spin dataset:} Scatter plot of the GW events used in the analysis. They correspond to the binary black hole GW events from GWTC-2.1 and GWTC-3 catalogs, selected with an $\rm IFAR\geq 1 yr$. The x-axis shows the source frame masses $\rm m^{\rm s}$ and the y-axis displays the dimensionless spin magnitudes $\chi$. The errors bars are the $1\sigma$ uncertainties of the official LVK parameter estimation samples for each GW event.}
    \label{fig:scatter plot mass source and chi}
\end{figure}

In this work we focus on the potential correlations between the BBH spin magnitudes and their masses. We study the GWTC-2.1 and GWTC-3 catalogs (https://zenodo.org/records/5546676) using new parametric population models to explore this interplay. Our analysis framework is validated on both simulated and real GW events, providing robust insights into the complex relationship between BBH mass and spin. 

\section*{\label{Sec:Results}Results}

We select a subset of 59 confident GW events with an inverse false alarm rate (IFAR) $\rm IFAR\geq 1 yr$. The estimated values of the spin magnitudes and source mass of these GW events are depicted in Fig.~\ref{fig:scatter plot mass source and chi}, with their respective errors from the parameter estimation samples provided by \cite{KAGRA:2021vkt}. Although visually the data suggests a correlation between the spin magnitude $\chi$ and the source frame mass of the BBHs $\rm m^{\rm s}$, we employ phenomenological models to try to model this correlation while also doing the deconvolution of the possible presence of selection biases.

\subsection*{\label{Sec:Models} Models of BBH populations}

To characterize the interplay between mass and spin, we construct three classes of parametric models that will be used for the hierarchical Bayesian inference. The analytical forms of the models, as well as their priors on the population parameters, are reported in the Supplement Material.

The first class is named \textsc{Evolving Gaussian} and describes the spin magnitude as a gaussian distribution with mean and variance that evolve linearly with the value of the source mass. The distribution of tilt angles follows the \textsc{Default} spin model in \cite{Wysocki:2018mpo, Talbot:2019okv, LIGOScientific:2020kqk}, where a fraction of the population has nearly aligned spins to the orbital angular momentum and the other isotropic. Spin's tilt distributions do not evolve with the mass of the model. For the primary mass of the system, we adopt a \textsc{PowerLaw + Peak} (PLP) model, while the secondary mass of the binary system is described by a Power Law (PL)~\cite{KAGRA:2021duu}.  We describe the binary merger rate based on the Madau \& Dickinson (MD) star formation rate \cite{Madau:2014bja}. 

The second model family introduces a mass transition between two populations with separate spin magnitude distributions. In one case, the two populations are described by a Beta and a gaussian distributions (\textsc{Beta to Gaussian}) while in the other, by two Beta distributions (\textsc{Beta to Beta}). The transition of spin distributions in mass is described by a logistic function, whose midpoint and steepness are free parameters. 

The third family of models is referred to as \textsc{Mixture} models. These models parameterize the overall population as the sum of two independent subpopulations \cite{Zevin:2017evb,Zevin:2020gbd}. The two subpopulations are combined using a mixture fraction. Each subpopulation has uncorrelated mass, redshift and spin distributions.
For all the \textsc{Mixture} models, the CBC merger rates are parameterized using a MD parameterization and the spins with the \textsc{default} spin model \cite{Wysocki:2018mpo, Talbot:2019okv, LIGOScientific:2020kqk}. We construct three \textsc{mixture} models. The \textsc{Mixture Vanilla} adopts a PLP and PL distributions for the primary masses of the first and second subpopulations. The \textsc{Mixture Paired} model uses the PLP and PL distributions to describe both the primary and secondary masses of the binaries \cite{Fishbach:2019bbm}. The \textsc{Mixture peak} describes the primary masses of the first population as a PL and the primary masses of the second population as a gaussian distribution. This last model is inspired from \cite{ray2024searching} that argues about the presence of a subpopulation of BBHs with different spin distribution in the excess of BBHs observed around 35 $M_\odot$.

\subsection*{\label{Sec:bayesfactor} Models Selection}

In Tab.~\ref{Tab:Bayes factor} we report the Bayes factors and the maximum of log-likelihood ratios, between a baseline population model for which the spins are not correlated with the masses and the models we introduced in the previous section. The baseline model employed for this work is a single population described by one PLP distribution for the primary mass and one \textsc{Default} model distribution for the spin, following ~\cite{KAGRA:2021duu}.

The Bayes factors reveal that all the models that parameterize the spin mass correlation as a transition between two subpopulations are strongly preferred against a model without any mass-spin correlation (despite the increased dimensionality of the fit). The \textsc{Evolving Gaussian} model that parameterizes the spin-mass relation as a continuous evolution, is not preferred nor excluded by the data.
\begin{table}[h]
    \centering
    \begin{tabular}{l c c} 
    \hline
    \textbf{Model} & \hspace{0.3cm} $\boldsymbol{\log_{10}\mathcal{B}}$ & \hspace{0.5cm} $\boldsymbol{\log_{10}\mathcal{L}_{\rm max}}$ \\ [0.9ex] 
    \hline 
    \textsc{Evolving Gaussian} & -0.48 & 2.94 \\ 
    \textsc{Beta to Gaussian} & 2.36 & 3.77 \\ 
    \textsc{Beta to Beta} & 2.55 & 3.91 \\ 
    \textsc{Mixture vanilla} & 2.78 & 4.66 \\ 
    \textsc{Mixture peak} & 1.64 & 1.25 \\ 
    \textsc{Mixture paired} & 3.78 & 5.99 \\ 
    \hline
    \end{tabular}
    \caption{Base 10 logarithm of the Bayes factors (second column) and the logarithm of the maximum likelihood ratio (third column), for the six models discussed in this letter compared to the reference model. The reference model is the canonical non evolving analysis (see Supplementary material).}
    \label{Tab:Bayes factor}
\end{table}
To further validate the Bayes factors, we performed several tests reported in the Supplement material. We have verified that the preference of the Bayes factor is due to the inclusion of the spin-mass relation, as analyses with only mass information are not able to discriminate between the models. We have verified that our spin-mass models can be confidently excluded when simulating a population of BBHs with no spin-mass correlation. Finally, we have further verified that Bayes factors are inconclusive when blinding real data to possible spin-mass correlation, this is done by shuffling spin/mass estimations for real GW data.

\subsection*{\label{Sec:Spinevol} The spin-mass correlation induced by a transition in mass between two subpopulations}

\begin{figure*}[h]
    \centering
    \includegraphics[width=\textwidth]{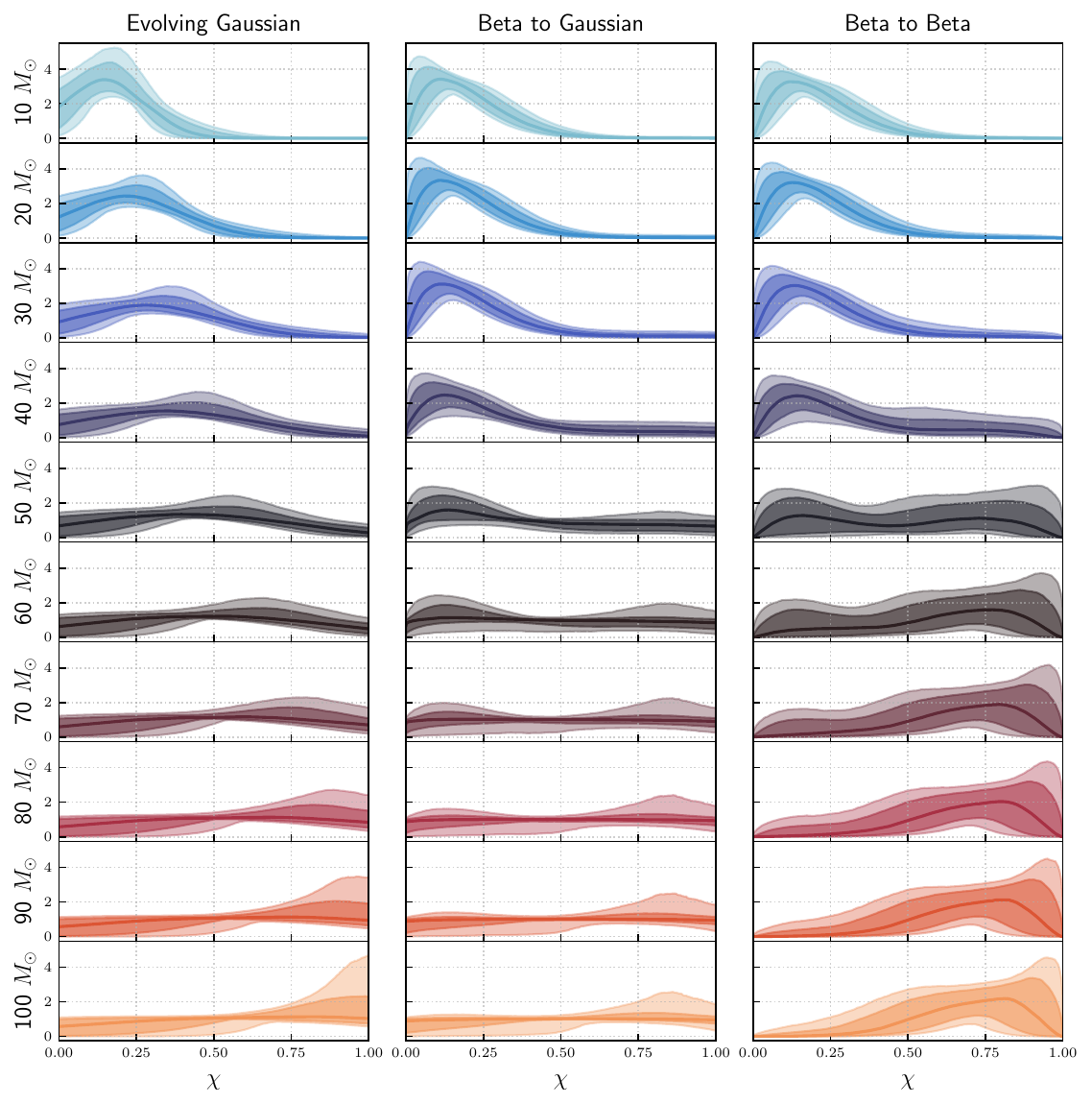}
    \caption{\textbf{Evolving spin magnitude:} This figure shows the probability density functions of the dimensionless spin magnitudes $\chi$, reconstructed from the full population inference on the 59 BBHs with an $\rm IFAR\geq 1 yr$ from the GWTC-2.1 and GWTC-3 catalogs, obtained with the \textit{Evolving Gaussian} (left column), \textit{Beta to Gaussian} (middle column) and the \textit{Beta to Beta} (right column) \textit{evolving} spin models. Each row represents a bin in source frame mass, from $10M_{\odot}$ up to $100M_{\odot}$, to highlight the spin-mass interplay.}
    \label{fig:spin magnitude joy plot evolving spin models}
\end{figure*}

Figure~\ref{fig:spin magnitude joy plot evolving spin models} shows the reconstructed spin distribution for the \textsc{Evolving gaussian}, \textsc{beta to gaussian} and \textsc{beta to beta} models. All the models reconstruct a transition from a population described by a low-spin magnitude distribution to a population described by a higher and wider spin magnitude distribution.  All three models infer a lowly spinning population (around $\chi \sim 0.2$) of compact objects at low masses transitioning around 40-50 $M_\odot$ to another population for which the spin distribution is surely higher (above 0.5) but can not significantly be constrained by the current data.

The \textsc{Evolving Gaussian} model describes the mass-spin evolution in two possible ways (see Supplement material) that can not be excluded from current data. Either the mean of the gaussian peak does not evolve with the mass, but the width of the distribution increases with it. Or the mean of the gaussian evolves with the mass and the width is nearly fixed. Both scenarios result in an evolving spin-mass distribution but can not yet be disentangled from current data.

The results from the (\textit{Beta to Beta} and \textit{Beta to Gaussian}) models that parameterized a transition between two subpopulations, see Fig.~\ref{fig:spin magnitude joy plot evolving spin models}, indicate a transition between two spin distributions between 40 and 55 $M_{\odot}$. From the current data, there is a preference for a steep transition ($\sim 10 M_\odot$) around 40 $M_{\odot}$, rather than a slightly wider ($\sim 20 M_\odot$) at 55 $M_{\odot}$, see Supplement material. This result indicates that the spin distribution of BHs at high masses is for sure higher than the one a low masses.

To understand if the spin-mass correlation is introduced by a wrong inference of the BBH mass spectrum, we compared the reconstructed mass distributions by our models with the ones of \cite{KAGRA:2021duu}. We have found that the mass spectrum reconstructed by our models is in excellent agreement with the ones inferred in \cite{KAGRA:2021duu} using an uncorrelated spin-mass model (see Supplement material).

\subsection*{\label{Sec:Subpop} The spin-mass correlation as the mixing of two independent subpopulations}

The results from the previous section support a spin-mass interplay induced by a transition happening around 40-55 $M_\odot$ between two populations with different spin distributions. With the \textsc{Mixture} models, we study if this relation could be consistent with the overlap of two independent subpopulations with separate and uncorrelated mass, spins and redshift distributions.

\begin{figure*}
    \centering
    \includegraphics[width=0.855\textwidth]{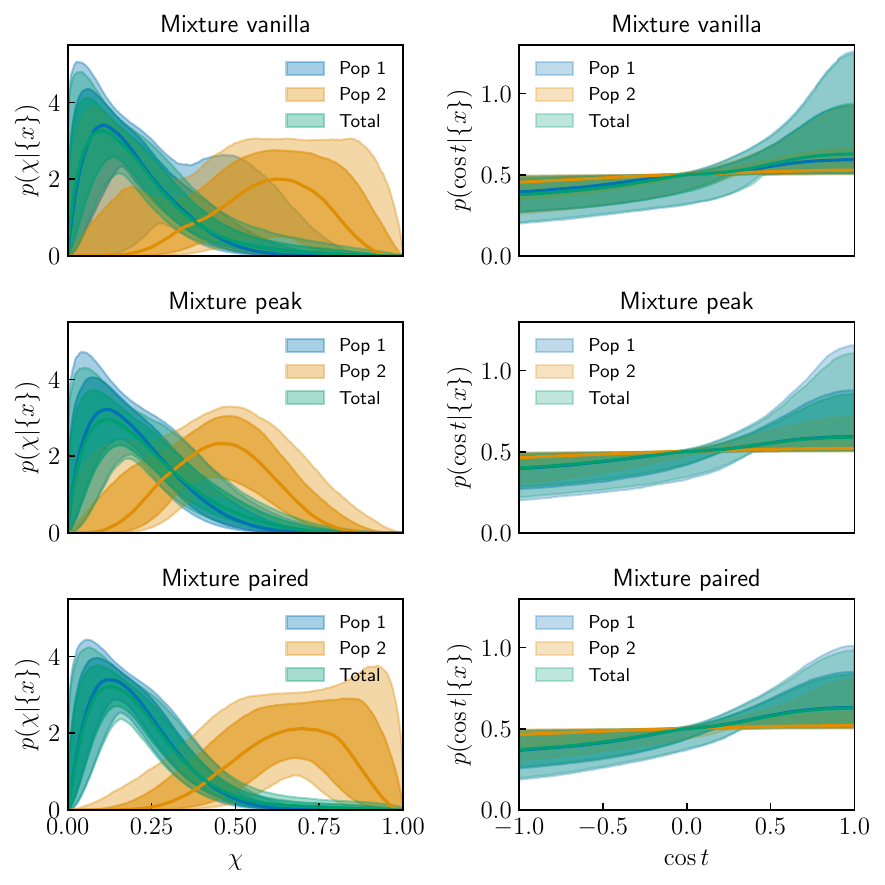}
    \caption{\textbf{Mixture spin distribution:} Reconstructed spectra of the spin magnitude $\chi$ and the cosine of the tilt angle for each of the three mixture analysis obtained on the 59 BBHs with an $\rm IFAR\geq 1 yr$ from the GWTC-2.1 and GWTC-3 catalogs. The blue curves (Pop1) are the spin magnitudes and tilt angles found for the first population and the yellow curves (Pop2) for the second population. The green curves (Total) are the combined distributions. The contours are the $90\%$ and $95\%$ C.L.}
    \label{fig:Spin magnitude mixture model PPC}
\end{figure*}

Fig.~\ref{fig:Spin magnitude mixture model PPC} depicts the reconstructed spin magnitudes and inclination angles inferred from each \textsc{mixture} model. For the \textsc{Mixture Vanilla} and \textsc{Mixture Paired} models, we set the primary population to describe the BBHs with masses $\lesssim 40-60 M_\odot$ (see prior ranges in Supplement material). For the \textsc{Mixture Peak}, the secondary population is described by a gaussian peak in the $20-50 M_\odot$ mass region. 

The common result among the \textsc{mixture} models is that the primary population of BBHs, which describes the large fraction of objects at small masses, supports low spin magnitudes peaking around $\chi \sim 0.1$, while the secondary population at higher masses supports a distribution of spins $\chi$ around 0.7. In all the cases, data supports the fact that first population accounts for almost 98\% of the overall population. See Supplement material for the posteriors on the mixture fraction of the population and its interplay with other population parameters. In terms of model selection, the data favors the secondary population with different spin magnitudes being located above $45 M_\odot$, rather than solely in the excess of BBHs around $35 M_\odot$.

For these models, we also reconstruct the spin's tilt angle distribution with respect to the orbital angular momentum. Interestingly, the inferred distributions for the primary (low-mass) population weakly prefer spins aligned with the orbital angular momentum, while the secondary (high-mass) population weakly prefers a more isotropic distribution. We note, however, that the reconstructions of the spins' tilt angle are still too uncertain to draw any robust conclusion about their distribution \cite{Vitale:2022dpa}.

For all the models, we verified that the reconstructed mass and redshift distributions are consistent with the  estimations obtained with non-evolving spin models and a single mass population \cite{KAGRA:2021duu} (see Supplementary material).

\section*{Discussion \label{sec4}}

Our analysis reveals novel compelling evidences for a correlation between the BBH spin magnitudes and their mass. This trend persists across all our models, and can either be described as a mass-dependent transition between two spin populations or the overlap of two independent subpopulations with uncorrelated spin, mass and redshift distributions. Moreover, all the models infer a lowly spinning population of BBHs at low masses and another population of BBHs at higher masses with a spin population that can only be loosely constrained.
We find the transition between the populations to happen at $40-55 M_\odot$. We find that the velocity and actual boundaries of the transition start to be constrained by the data.

Previous studies have already tried to inspect a possible mass-spin relation at the population level of BBHs \cite{Callister:2021fpo, Biscoveanu:2022qac,  Fishbach:2022lzq,Godfrey:2023oxb, Li:2023yyt}. In \cite{KAGRA:2021duu, Tiwari:2021yvr}, it is argued that the absolute value of the \textit{spin projection} over the orbital angular momentum does not significantly evolve with the chirp mass. This result is not in contrast with our findings (see supplement material), as the spin tilt angles are poorly constrained in terms of masses. Therefore the spin projection over the orbital angular momentum shows no particular correlation with the mass, although the spin magnitude can display a significant correlation.
Our results from the \textsc{Mixture peak} model are also consistent with the findings of \cite{ray2024searching} that, using a binned non-parametric model, argues for the presence of a subpopulation of BBHs with different spin distribution in the mass range $30-40 M_\odot$. Differently from our work, \cite{ray2024searching} focuses on the effective and precession spin parameters, which are a combination of spin magnitudes, tilt angles and masses, and uses a non-parametric binned model.

The results presented in \cite{Kimball:2020qyd} argue that the BBHs present in GWTC-2 support a population composed of $1^{st}-1^{st}$ generation, $1^{st}-2^{nd}$ generation and $2^{nd}-2^{nd}$ generation BHs. In \cite{Kimball:2020qyd}, the mass and spin distribution of BBHs for $1^{st}-1^{st}$ generation binaries are fit using phenomenological models like the ones employed in this paper, while the $1^{nd}-2^{nd}$ and $2^{nd}-2^{nd}$ mass and spin distributions are obtained with transfer functions defined in \cite{Kimball:2020opk} from the $1^{st}-1^{st}$ generation binaries. In our study, we go beyond the use of a transfer function calibrated on hierarchical formation channels only and, using general phenomenological models, we demonstrate that there is a spin-mass relation that is possibly introduced by the transition in mass between subpopulations described by different spin distributions. Nevertheless, the general results that we obtain are not in contrast with the conclusions of \cite{Kimball:2020qyd} that argues about the possible presence of $1^{st}-2^{nd}$ and $2^{nd}-2^{nd}$ generation binaries subpopulations.

In \cite{Li:2022gly, Li:2023yyt}, the investigation focuses on subpopulations of BBHs utilizing a semi-parametric approach. They observe hints of a transition in mass between two spin distributions while keeping the CBC merger rate constant and employing a single model for the tilt angle across both populations. Their approach resembles our \textsc{mixture vanilla} model, however it may not fully capture the complexity of the data- constraining the merger rate and the tilt angles in this manner could limit the nuanced understanding of BBH formations origins. Furthermore, our approach extends beyond theirs by exploring a wider range of scenarios and allowing for greater flexibility, thus providing a more comprehensive analysis of BBH subpopulations.

We now examine the possible astrophysical implications of our results regarding the BBHs formation channels. One of the most accepted theories for compact objects' formation is that BHs from isolated stellar binaries can not be formed beyond $[45-60]M_{\odot}$. This mass scale is identified as the lower edge of the Pair Instability Supernova (PISN) \cite{Farmer:2019jed,Farmer:2020xne, Renzo:2020rzx,Karathanasis:2022rtr}. In this picture, the PISN mass scale would mark a transition between a population of $1^{st}$-generation BHs formed by their stellar progenitors to a population of $n^{th}$-generation BHs dynamically assembled into binaries \cite{Mapelli:2021taw,Kimball:2020qyd} in dense stellar environments. The population of $1^{st}$-generation BBHs is predicted to have relatively small spins aligned to the orbital angular momentum \cite{Mapelli:2021taw}, due to the various astrophysical processes expected to happen during the stellar binary evolution. The population of  $n^{th}$-generation BHs' is expected to display spins magnitude around 0.7 (from the pre-merger binary) and nearly isotropically distributed \cite{Berti:2008af,Gerosa:2017kvu,Fishbach:2017dwv,GalvezGhersi:2020fvh}. According to the latest BBH synthesis simulations, $1^{st}$ generation BBHs are expected to form 97.5-98\% of the population while $n^{th}$-generation the rest, all formation channels combined \cite{Li:2022gly,Mapelli:2021syv}. 
In this picture, we do not include the contribution of BHs formed by population III stars, which are likely to be located at very high redshifts (not accessible by current data) and compose a very small fraction of the overall astrophysical population.

Our results seem to support this scenario. The transition between the subpopulations that we observe around 40-55 $M_\odot$ could be linked to the PISN gap. The lower-mass population displays a clear preference for low spin magnitudes ($\chi \sim 0.1$) as expected from BHs formed in isolated stellar binaries. Although, the spin distribution for the higher mass population is not as strongly constrained as the low mass one, we know that this is surely different from the spin distribution of the low-mass population, and we find new evidence from the \textsc{mixture} models that the spin distribution supports values around 0.7. Moreover, the \textsc{Mixture} models infer that the high-mass ($n^{th}$-generation) population should compose only the $2\%$ of the astrophysical BBHs.

Another relevant result found by the \textsc{Mixture} models is that the BBH merger rate as a function of redshift increases in the same way for the low and high mass subpopulations (see Supplement material). If we identify the former population as $1^{st}$-generation BHs and the second with $n^{th}$ generation BHs, this result would imply that the time scales over which hierarchical mergers happen are cosmologically small.

The correlations we observed between the spin magnitudes and mass support the existence of two subpopulations, transitioning around 40-55 $M_\odot$, described by different spin distributions. These findings provide support for the existence of $n^{th}$ generation BBH mergers from the hierarchical merger formation channel. However, definitive evidence for this hypothesis could be reached with a better reconstruction of the spin tilt distribution that could be obtained with future GW observations. 

\section*{\label{Sec:Method} Method}
We estimate the parameters governing the population properties of BBHs, including their masses, CBC merger rates, and spins, based on a set of detected GW events. This analysis is conducted within a hierarchical Bayesian inference framework, where astrophysically motivated population models are inferred. Specifically, we employ \icarogw, a code developed for inferring population properties from noisy and heterogeneous GW data while accounting for selection effects \cite{Mastrogiovanni:2023zbw, Mastrogiovanni:2023emh}.

To estimate selection effects within the Bayesian framework, we utilize the public LVK set of detected injections, covering the entire parameter space of interest (https://doi.org/10.5281/zenodo.7890398) \cite{KAGRA:2021duu, KAGRA:2023pio}. We ensured numerical stability by utilizing a sufficient number of injections (see Supplementary Material). Additionally, given the focus of this work on the interplay between mass and spin, we fix the cosmological parameters ($H_{0},\Omega_{\rm m}^{0}$) to the Planck 2015 measurement during population inference \cite{Planck:2015fie}.

\begin{acknowledgments}
The authors are grateful for computational resources provided by the LIGO Laboratory and supported by the National Science Foundation Grants PHY-0757058 and PHY-0823459. This material is based upon work supported by NSF's LIGO Laboratory which is a major facility fully funded by the National Science Foundation.

\end{acknowledgments}

\newpage
\clearpage

\bibliographystyle{unsrt}
\bibliography{apssamp}

\newpage
\clearpage
\onecolumngrid
\begin{center}
    \Large\bfseries Supplementary Material and Supplementary Figures
\end{center}

\vspace{2cm}
\twocolumngrid

\section{Hierarchical Bayesian inference}

The detection of Gravitational Wave (GWs) sources by current ground-based detectors can be described as an inhomogeneous Poisson process in the presence of selection biases. The central quantity of inference is the Binary Black Hole merger rate 
\begin{equation}
     \frac{\de \Ncbc}{\de \theta \de z \de t_s}(\Lambda)
     \label{eq:rate}
\end{equation}
that we describe in terms of binary parameters $\theta$ (in this work, these are the source masses, spin magnitudes, tilt angles, CBC rate), redshift and source time. The BBH rate in Eq.~\ref{eq:rate} is function of population parameters $\Lambda$, the rate models we employ in this work are described in Sec.~\ref{sec:models}. The hyperlikelihood can be written as \cite{Vitale:2020aaz,MANDEL20221}
\begin{eqnarray}
    \mathcal{L}(\{x\}|\Lambda) &\propto&  e^{-N_{\rm exp}(\Lambda)} \prod_i^{\Ngw} \Tobs \int \de \theta  \de z \; \mathcal{L}_{\rm GW}(x_i|\theta,z)  \nn \\  && \times \frac{1}{1+z} \frac{\de \Ncbc}{\de \theta \de z \de t_s}(\Lambda).
    \label{eq: hyperlikelihood}
\end{eqnarray}
In the above Eq.~\ref{eq: hyperlikelihood}, $\mathcal{L}_{\rm GW}(x_i|\theta,z)$ is the GW likelihood, it quantifies the errors on the estimation of the $\theta$ and $z$ parameters from the data.
The term \Nexp takes into account the selection effects, due to the finite sensitivity of the detectors,
\begin{equation}
    \Nexp (\Lambda)= \Tobs \int \de \theta \de z \; \Pdet(\theta, z) \, \frac{1}{1+z} \frac{\de \Ncbc}{\de z  \de \theta \de t_s}(\Lambda) ,
    \label{eq:Nexp}
\end{equation}
where the detection probability $\Pdet(z,\theta)$ represents the probability that an event characterized by its true binary parameters $\theta$ at redshift $z$ is detected, i.e. it overpasses some chosen detection threshold adopted by the search algorithm (e.g. the SNR or the false alarm rate, FAR). 
 
The hierarchical likelihood is evaluated numerically for each population model using a set of parameter estimation samples from \Ngw GW events and a set of detectable injections that are used to evaluate selection biases. Each injection and parameter estimation sample consists in a value for the parameters $\theta$ and $z$, that is used to evaluate the BBH merger rate in Eq.~\ref{eq:rate} \cite{Mastrogiovanni:2023zbw} and to deconvolve the priors $\pi_{\rm PE}$ and $\pi_{\rm inj}$ that are used to generate the parameter estimation samples and the injections. The overall likelihood is approximated as
\begin{equation}
    \ln[\mathcal{L}(\{x\}|\Lambda)] \approx -\frac{T_{\rm obs}}{N_{\rm gen}} \sum_{j=1}^{N_{\rm det}} s_j + \sum_{i}^{N_{\rm obs}} \ln\left[ \frac{T_{\rm obs}}{N_{{\rm s},i}} \sum_{j=1}^{N_{{\rm s},i}} w_{i,j} \right]\,,
        \label{eq:hl_numer}
\end{equation}
where $s_j$ and $w_{i,j}$ are the weights associated to the injections and parameter estimation samples, the $i$ index refers to the $i$-th event, while $j$ to the Monte Carlo sampling:
\begin{eqnarray}
    s_j & = & \frac{1}{\pi_{\rm inj}(\theta_j, z_j)}\frac{1}{1+z_j}\frac{dN}{dt_s dz d\theta}(\Lambda)\bigg|_j \\
    w_{i,j} & = & \frac{1}{\pi_{\rm PE}(\theta_{i,j}, z_{i,j}|\Lambda)}\frac{1}{1+z_{i,j}}\frac{dN}{\de t_s \de \theta \de z}(\Lambda)\bigg|_{i,j}
\end{eqnarray}

In our analyses, we apply a set of criteria to check the numerical stability of Eq.~\ref{eq:hl_numer}. In particular, we require that at least 10 parameter estimation samples per GW event contribute to the numerical evaluation of the likelihood and at least 200 injections for the calculation of the selection bias \cite{Farr:2019rap}.

\section{Population models and priors}
\label{sec:models}

In our analysis, we use three classes of population models to parameterize the BBH merger rate given in Eq.~\ref{eq:rate}. The first class has only one population, the \textsc{vanilla} model, that describes the spin, redshift and mass distributions and independent from each other and uncorrelated. The second class, \textsc{evolving models}, includes three population models that describe the mass and spin distribution as independent while modelling the spin distribution with an analytical dependence from the mass. The third class, \textsc{Mixture models}, describe the overall BBH merger rate as the superposition of two independent sub-populations with correlated mass, spins and redshift distributions.

\subsection{Vanilla model}

The BBH rate function is described as 
\begin{equation}
    \frac{\de \Ncbc}{\de \theta \de z \de t_s}(\Lambda) = R(z;\Lambda) \frac{\de V_c}{\de z}p_{\rm pop}(\vec{m}_s|\Lambda) \pi(\vec{\chi_1},\cos \vec{\theta}|\Lambda),
\end{equation}
where the vectors indicate the components of the two binary masses, spin magnitudes and tilt angles. The priors used for this run are listed in Tab.~\ref{tab: Vanilla analysis}.

The rate function is modelled after a Madau-Dickinson (MD) \cite{Madau:2014bja} star formation rate s.t.
\begin{equation}
    R(z;\Lambda)=R_0 [1+(1+z_p)^{-\gamma-k}] \frac{(1+z)^\gamma}{1+\left(\frac{1+z}{1+z_p}\right)^{\gamma+k}} \,.
    \label{eq:ratemod2}
\end{equation}
The mass distribution is modeled according to a \textsc{Power Law + peak} model with
\begin{eqnarray}
    \pi(m_{1,s}|m_{\rm min},m_{\rm max},\alpha)&=&(1-\lambda)\mathcal{P}(m_{1,s}|m_{\rm min},m_{\rm max},-\alpha)+ \nonumber \\ && \lambda \mathcal{G}(m_{1,s}|\mu_g,\sigma)\,, \quad (0 \leq \lambda\leq 1) \label{eq:pl1}\\
    \pi(m_{2,s}|m_{\rm min},m_{1,s},\beta)&=&\mathcal{P}(m_{2,s}|m_{\rm min},m_{1,s},\beta)\,.
    \label{eq:pl2}
\end{eqnarray}
where $\mathcal{P}$ is a truncated powerlaw and $\mathcal{G}$ a gaussian. The mass distribution also include a tapering at low masses governed by a population parameter $\delta_m$.
The spin distribution is modelled according to the \textsc{Default} spin model of \cite{Talbot:2019okv,PhysRevD.100.043012} as
\begin{eqnarray}
    \pi(\vec{\chi},\cos \vec{\theta}|\Lambda) &=& {\rm Beta}(\chi_1|\alpha,\beta) {\rm Beta}(\chi_2|\alpha,\beta) \times \nonumber \\ && \pi(\cos \vec{\theta}|\xi,\sigma_t) ,
\end{eqnarray}
with 
\begin{equation}
    \pi(\cos \theta_{1,2}|\zeta,\sigma_t)=\xi \mathcal{G}_{[-1,1]}(\cos \theta_{1,2}|1,\sigma_t) +\frac{1-\xi}{2},
    \label{eq:B47}
\end{equation}
where $\mathcal{G}_{[-1,1]}(\cos \theta_i|1,\sigma_t)$ is a  truncated Gaussian between $-1$ and $1$. 

\subsection{Evolving models}

The BBH rate function is described as 
\begin{equation}
    \frac{\de \Ncbc}{\de \theta \de z \de t_s}(\Lambda) = R(z;\Lambda) \frac{\de V_c}{\de z}p_{\rm pop}(\vec{m}_s|\Lambda) \pi(\vec{\chi},\cos \vec{\theta}|\vec{m},\Lambda),
\end{equation}
Differently from the vanilla case, the spin for this class of models is conditioned on the value of the source mass. For all the models, the spin distribution is factorized as 
\begin{equation}
\pi(\vec{\chi},\cos \vec{\theta}|\vec{m},\Lambda) = \pi(\vec{\chi}|\vec{m},\Lambda) \pi(\cos \vec{\theta}|,\Lambda),  
\end{equation}
where the spin magnitude probability is conditioned on the mass values (see below for the models) and the angular distribution $\pi(\cos \vec{\theta}|,\Lambda)$ is not dependent on the mass and it is the one for the \textsc{default} model. Moreover, all the models assume that
\begin{equation}
\pi(\vec{\chi}|\vec{m},\Lambda) = \pi(\chi_1|m_1,\Lambda) \pi(\chi_2|m_2,\Lambda).
\end{equation}
For all the \textsc{Evolving} models, we use the rate model in Eq.~\ref{eq:ratemod2} and the mass spectrum described by the \textsc{power law + peak} in Eq.~\ref{eq:pl1} and Eq.~\ref{eq:pl2}. The priors used for this run are listed in \cref{tab: Mass Evolving and shifting,tab: Rate Evolving models,tab: Spin Evolving model,tab: Spin Beta to Gaussian model,tab: Spin Beta to Beta model}. 
Below we describe what parametrization is used for the spin magnitude distribution.

\subsubsection{Evolving gaussian}

The \textsc{evolving} gaussian model parameterizes the spin magnitudes distribution as a truncated gaussian between 0 and 1 with a mass-varying mean and standard deviation, namely
\begin{equation}
    \pi(\chi|m,\Lambda) = \mathcal{G}_{[0,1]}(\chi|\mu(m),\sigma(m)).
\end{equation}
The mass-varying mean and standard deviation are approximated at the linear with a first order Taylor's expansion.
\begin{eqnarray}
\mu(m) &=& \mu_0 + \dot{\mu}m  \\   
\sigma(m) &=& \sigma_0 + \dot{\sigma}m.
\end{eqnarray}
For this model, $\mu_0, \sigma_0, \dot{\mu}, \dot{\sigma}$ are additional population parameters

\subsubsection{Beta to Gaussian}

This model parameterizes the spin distribution as a mass-dependent transition from a spin population described by a $Beta(\chi|\alpha,\beta)$ distribution to a spin population described by a truncated gaussian $\mathcal{G}_{[0,1]}(\chi|\mu,\sigma)$. The spin distribution is given by
\begin{eqnarray}
    \pi(\chi|m,\Lambda) &=& W(z;m_{\rm t},\delta m_{\rm t}) B(\chi|\alpha,\beta) \nn \\ &&+  (1-W(z;m_{\rm t},\delta m_{\rm t})) \mathcal{G}_{[0,1]}(\chi|\mu,\sigma),
\end{eqnarray}
where $W(z;m_{\rm t},\delta m_{\rm t})$ is a logistic function that smoothly transition from 1 to 0. The window function is defined as 
\begin{equation}
    W(z;m_{\rm t},\delta m_{\rm t}) = \frac{1}{1 + e^{\frac{m-m_{\rm t}}{\delta m_{\rm t}}}}.
\end{equation}

\subsubsection{Beta to Beta}

This model is similar to the \textsc{beta to gaussian} but instead, it parameterizes the spin function as transitioning between two Beta distributions, namely
\begin{eqnarray}
    \pi(\chi|m,\Lambda) &=& W(z;m_{\rm t},\delta m_{\rm t}) B(\chi|\alpha_1,\beta_1) \nn \\ &&+  (1-W(z;m_{\rm t},\delta m_{\rm t})) B(\chi|\alpha_2,\beta_2).
\end{eqnarray}
The window function is defined as in the previous case.

\subsection{Mixture models}

This class of models parametrized the BBH merger rate as the overlap of two sub-populations, population 1 (Pop1) and Population 2 (Pop2). The CBC merger rate is given by 
\begin{equation}
    \frac{\de \Ncbc}{\de \theta \de z \de t_s}= \lambda_{\rm pop}\frac{\de \Ncbc}{\de \theta \de z \de t_s}(\Lambda^{\rm Pop 1}) + (1-\lambda_{\rm pop})\frac{\de \Ncbc}{\de \theta \de z \de t_s}(\Lambda^{\rm Pop 2}),
\end{equation}
where the parameter $\lambda_{\rm pop}$ parameterized the mixture fraction between the two subpopulations. The parameters $\Lambda^{\rm Pop 1}, \Lambda^{\rm Pop 2}$ collectively indicate the population parameters of the two sub-populations that are independent from each other.
The \textsc{mixture} models parameterize the sub-population merger rates as with distributions of mass, spins and redshift that are uncorrelated. The specific factorization of the sub-population merger rates is described in the section below and the priors and population parameters used are indicated in \cref{tab: Mass Mixture vanilla model,tab: Mass Mixture peak model,tab: Rate Mixture vanilla model,tab: Spin Mixture vanilla model}. 

\subsubsection{Mixture vanilla}

The \textsc{mixture vanilla} model parameterizes the distribution of the first population using a \textsc{power law + peak} for the primary and secondary mass distribution as in Eqs~\ref{eq:pl1}-\ref{eq:pl2}, a \textsc{default} spin model for the spin magnitudes and orientation a MD rate function for the BBH merger rate.
The second population is modelled with a mass spectrum described for the primary mass distributed according to a truncated power law, the secondary mass has a distribution as the one in Eq.~\ref{eq:pl2}. The spin magnitudes and orientation according to a \textsc{default} model and a MD rate function for the BBH merger rate.

\subsubsection{Mixture peak}

The \textsc{mixture peak} model parameterizes the distribution of the first population using a power law for the primary mass and secondary mass distribution as in Eq.~\ref{eq:pl2}, a \textsc{default}. The spin magnitudes and orientation are distributed according to a \textsc{default} model and the BBH rate function as a MD rate.
The second population is modelled with a mass spectrum described by the primary and secondary masses distributed according to a gaussian with mean $\mu_g$ and standard deviation $\mu_g$. An additional constraint is set on the gaussian mass distribution to ensure that $m_1>m_2$. The spin magnitudes and orientation according to a \textsc{default} model and a MD rate function for the BBH merger rate.

\subsubsection{Mixture Pairing}

The \textsc{mixture pairing} model is a variation of the \textsc{mixture vanilla} model. The distributions of spins, redshift and primary masses of the two sub-populations are still described as in the \textsc{mixture vanilla} model, with the difference that the secondary mass is forced to have the same distribution as the primary mass. In other words, the mass distribution is generally factorized as  
\begin{equation}
    p_{\rm pop}(m_1,m_2|\Lambda) \propto p_{\rm pop}(m_1|\Lambda) p_{\rm pop}(m_2|\Lambda) \left[\frac{m_2}{m_1}\right]^\beta \Theta(m_1-m_2), 
\end{equation}
where $\Theta$ is an Heaviside step function forcing $m_1>m_2$ and an additional weight dependent on the two masses ratio is included.

\section{Additional material for results}

\subsection{Evolving population models}

From the results obtained with the \textsc{evolving} models, we observe a consistent evolution of the spin magnitude distribution across the mass range. This evolution progresses from lower spins at lower masses to higher spins at higher masses. However, the higher-mass distribution of spin magnitudes is not as tightly constrained as that of lower masses.

\begin{figure*}
    \centering
    \includegraphics[width=\textwidth]{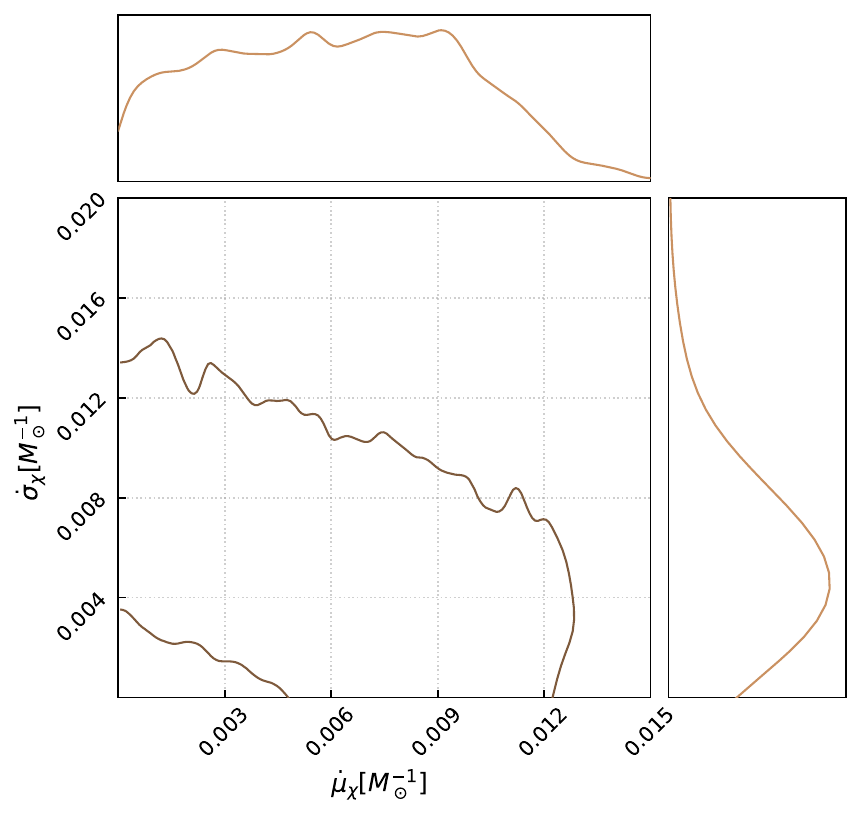}
    \caption{\textbf{Evolving Gaussian parameters:} Corner plot of the population parameters $\dot{\mu}_{\chi}$ and $\dot{\sigma}_{\chi}$ of the \textsc{Evolving Gaussian} model, obtained from the population inference of 59 real GW events from GWTC-2.1  GWTC-3 catalogs.}
    \label{fig:evolution_gaussianreal_ana}
\end{figure*}

With the \textsc{evolving gaussian} model, two potential spin evolution scenarios emerge, as illustrated in Fig.~\ref{fig:evolution_gaussianreal_ana}. Either the spin distribution maintains a stable peak position while spreading out towards higher spin magnitudes, or the width of the Gaussian distribution remains relatively fixed while the position of the distribution shifts. In either case, the model indicates an evolution in spin magnitudes across mass ranges.

\begin{figure*}
    \centering
    \includegraphics[width=\textwidth]{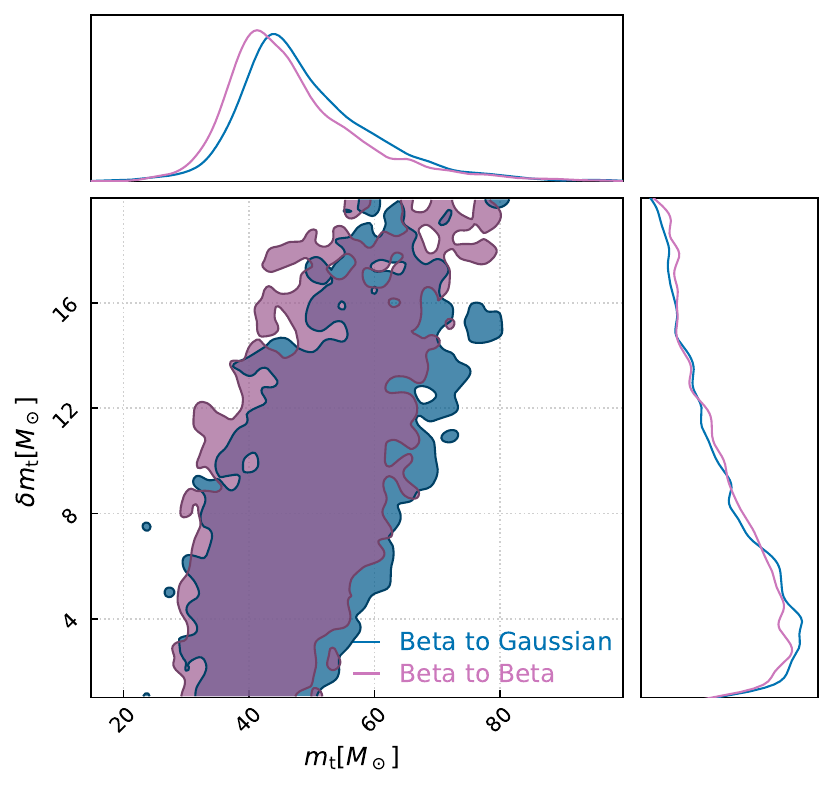}
    \caption{\textbf{Critical Mass transition-steepness:} Corner plot of the mass transition point $\rm m_{t}$ and the transition steepness $\delta_{m_{t}}$ population parameters, obtained from the population inference of 59 real GW events from GWTC-2.1  GWTC-3 catalogs with the \textsc{Evolving} models with a window function.}
    \label{fig:transitionreal_ana}
\end{figure*}

\begin{figure*}
    \centering
    \includegraphics[width=\textwidth]{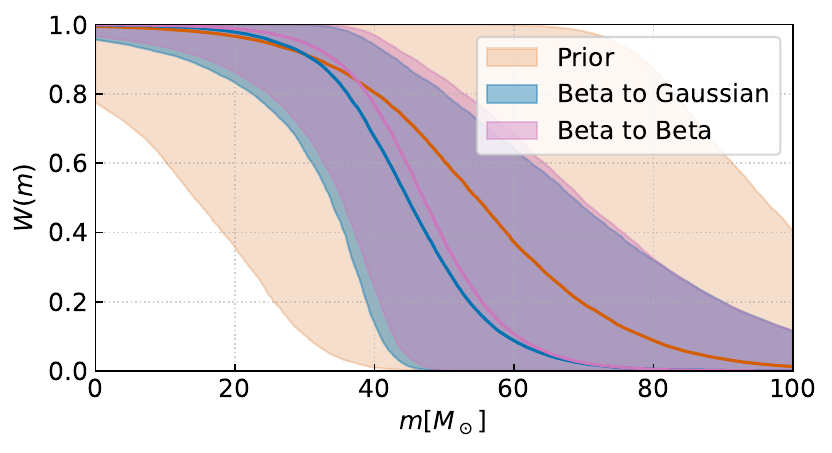}
    \caption{\textbf{Window function:} Reconstruction of the window function $W(m)$ as a function of the mass, obtained from the population inference of GWTC-3 data with the \textsc{Evolving} models. The colored contours correspond to the $90\%$ C.L.}
    \label{fig:windowreal_ana}
\end{figure*}

The two \textsc{evolving} models incorporating a window function, namely the \textsc{Beta to Beta} and \textsc{Beta to Gaussian} models, suggest the possibility of a transition from one spin magnitude distribution to another, that is preferred against the \textsc{Evolving gaussian} model. These models support either a rapid transition around 40 $M_{\odot}$ or a smoother transition at higher masses (60 $M_{\odot}$), as depicted in Fig.~\ref{fig:transitionreal_ana} and Fig.~\ref{fig:windowreal_ana}. Based on the population inference of GWTC-3 events using the window models, neither scenario can be excluded definitively. However, they all indicate a transition in spin magnitudes around 40-60 $M_{\odot}$ from slower to more rapid spinning objects.

The angular spin reconstruction of the tilt angles, the CBC merger rate reconstruction, and the sources' frame mass reconstruction all align closely with each other, particularly with the non-evolving baseline analysis, as demonstrated in Fig.~\ref{fig:mass_reconstructionreal_ana}.

\begin{figure*}
    \centering
    \includegraphics[width=0.7\textwidth]{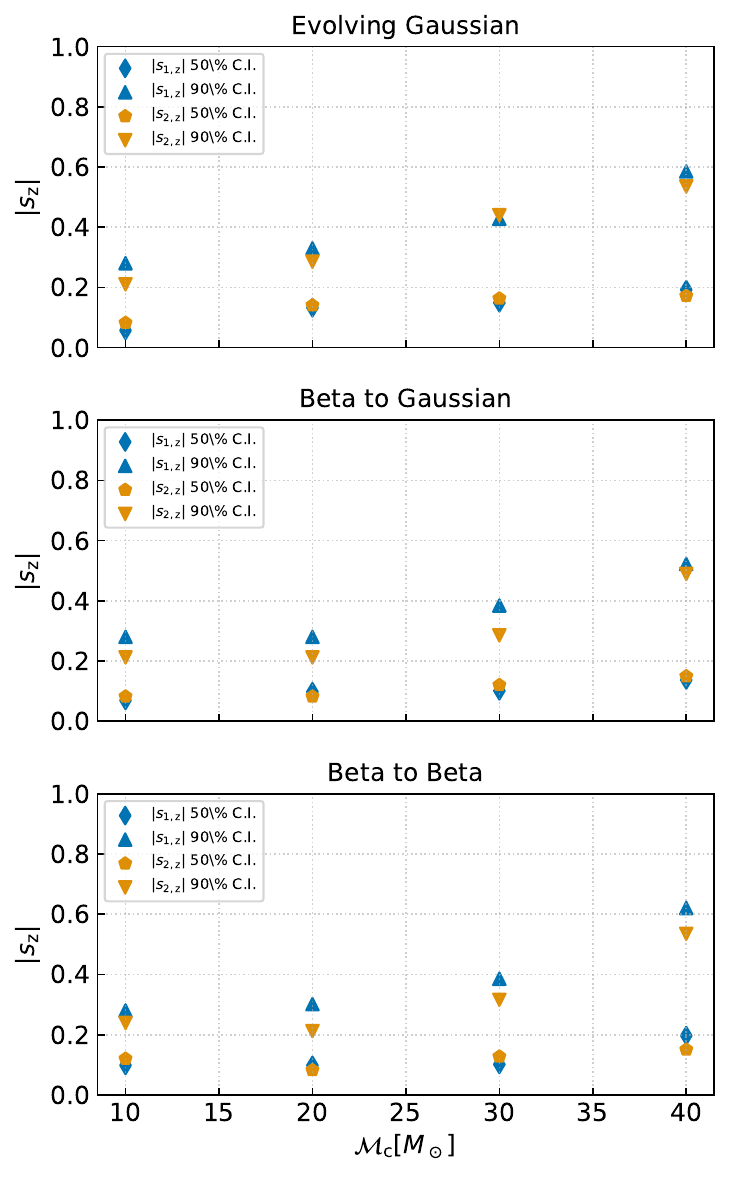}
    \caption{\textbf{Aligned spin magnitude evolution:} Scatter plot representing the evolution of the aligned component of the spin magnitude $\rm s_{z}$ with respect to the chirp mass $\mathcal{M}_c$,obtained from the population inference of 59 real GW events from GWTC-2.1  GWTC-3 catalogs, obtained using the \textsc{Evolving Gaussian} (top), the \textsc{Beta to Gaussian} (middle) and \textsc{Beta to Beta} (bottom) models.}
    \label{fig:chirp mass spin}
\end{figure*}

We performed an extra test, following \cite{KAGRA:2021duu}, we compare the reconstructed distribution of the spin magnitudes aligned with the orbital angular momentum, $|s_{z}|$, as a function of detector chirp mass $\mathcal{M}_{c}$. Fig.~\ref{fig:chirp mass spin} shows the reconstructed 50\% and 90\% credible upper bounds for the spin magnitude aligned with the orbital angular momentum. Upper bounds seem to evolve to more rapidly spinning BHs starting around 40 $M_{\odot}$.  While in \cite{KAGRA:2021duu} this trend is explained as due to a lower constraint on the spin for massive BHs, here we still obtain as a consequence of a spin-mass relation (this trend is not observed if we remove the spin-mass relation, see Sec.~\ref{sec:sanity}). 

\subsection{Mixture population models}

\begin{figure*}
    \centering
    \includegraphics[width=\textwidth]{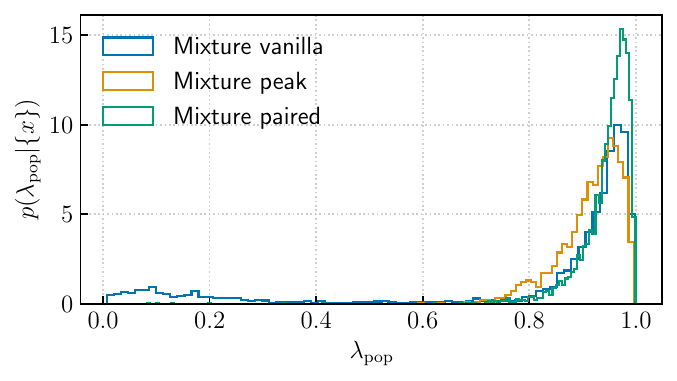}
    \caption{\textbf{Mixture parameter:} Histograms of the inferred posterior distribution of the mixture parameter $\lambda_{\rm pop}$ for three different flavours of the \textsc{Mixture} model analysis on 59 real GW events from GWTC-2.1  GWTC-3 catalogs, namely the \textsc{Mixture vanilla} (blue), \textsc{Mixture peak} (orange) and \textsc{Mixture} paired (green).}
    \label{fig:mixture_posteriorreal_ana}
\end{figure*}

\begin{figure*}
    \centering
    \includegraphics[width=\textwidth]{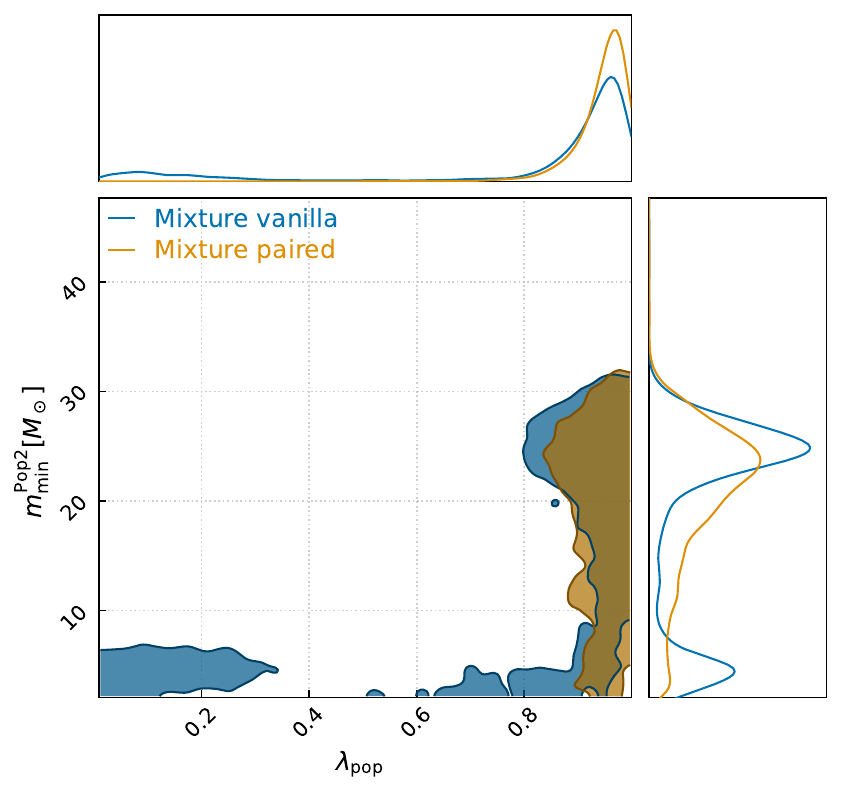}
    \caption{\textbf{Mixture bi-modality:} Corner plot of the mixture parameter $\lambda_{\rm pop}$ and the minimum mass of the second population $\rm m_{\rm min}^{\rm pop2}$ inferred with the \textsc{Mixture} vanilla and \textsc{Mixture paired} model from the population inference of 59 real GW events from GWTC-2.1  GWTC-3 catalogs. The contours are the $90\%$ C.L estimated.}
    \label{fig:double_peak_real_ana}
\end{figure*}

In this section, we present further explanations of the results obtained with the \textsc{mixture} models. In particular, when the second population is allowed to arrive down to 2 $M_{\odot}$, the posterior distribution of the mixing parameter $\lambda_{\rm pop}$ exhibits a slight bi-modality with the minimum mass of the second distribution ($\rm m_{\rm min}^{\rm pop2}$). This bi-modality is presented in Fig.~\ref{fig:mixture_posteriorreal_ana} and Fig.~\ref{fig:double_peak_real_ana}, and suggests that there is some support for a lower value of $\lambda_{\rm pop}$, finding at the same time a minimum mass for the second population close to a few solar masses. This minimum mass corresponds to the one of the first population. Nonetheless, the mode given by $\lambda_{\rm pop}\sim 0.98$ and $\rm m_{\rm min}^{\rm pop2}\sim 35 M_{\odot}$ is strongly preferred by the analysis, thus favouring the hypothesis that a sub-population of BHs with different spins could begin around $35 M_\odot$.

\begin{figure*}
    \centering
    \includegraphics[width=0.6\textwidth]{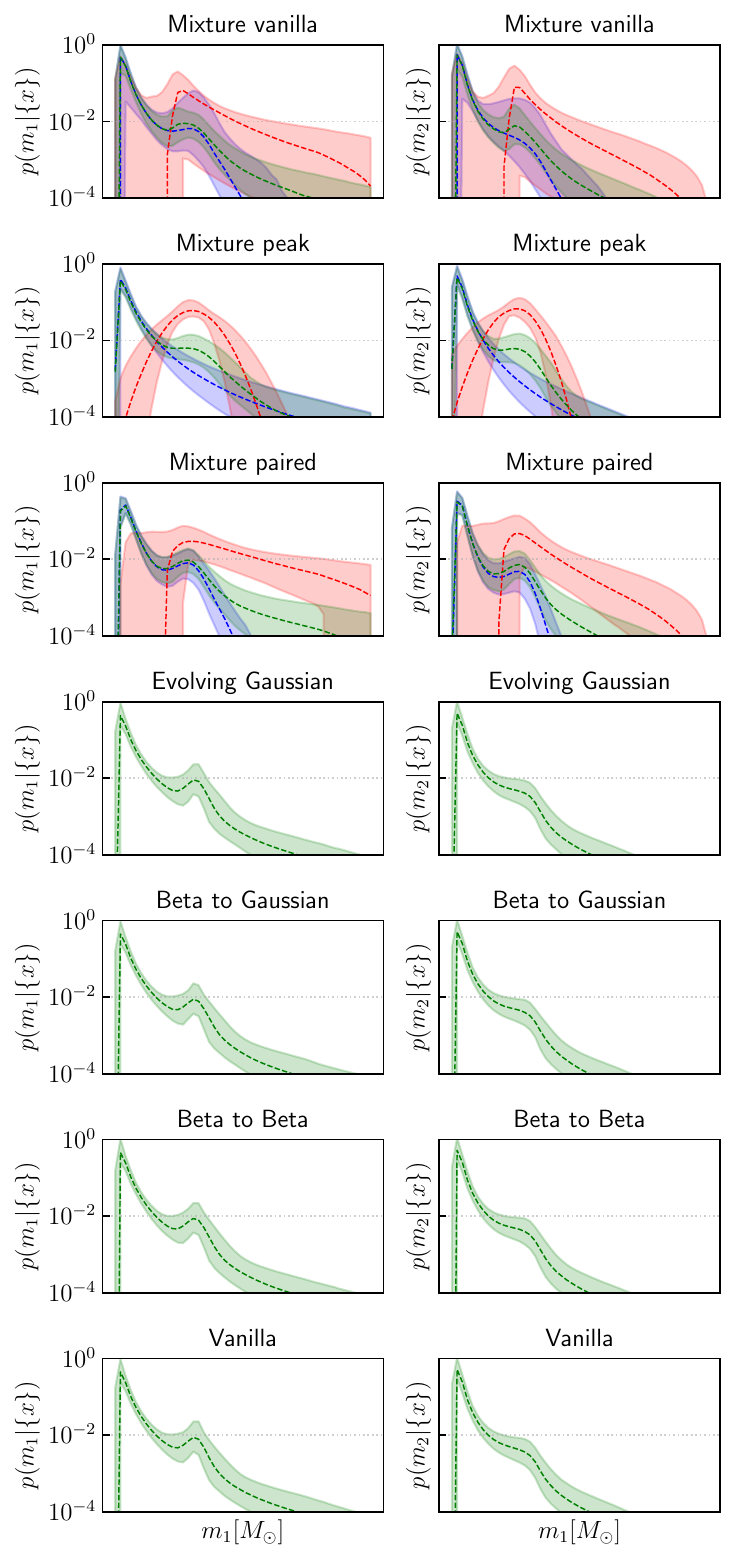}
    \caption{\textbf{Mass spectrum reconstruction:} Reconstructed mass spectra of the primary (left column) and secondary (right column) masses obtained from the population inference of 59 real GW events from GWTC-2.1  GWTC-3 catalogs, with the \textsc{Mixture vanilla} model (first row), \textsc{Mixture peak} (second row), \textsc{Mixture paired} (third row), \textsc{Evolving gaussian} (fourth row), \textsc{Beta to Gaussian} (fifth row), \textsc{Beta to Beta} (sixth row) and the canonical \textsc{Vanilla} model (seventh row). The dotted lines are the median values and the colored contours the $90\%$ C.L. inferred.}
    \label{fig:mass_reconstructionreal_ana}
\end{figure*}

\begin{figure*}
    \centering
    \includegraphics[width=0.7\textwidth]{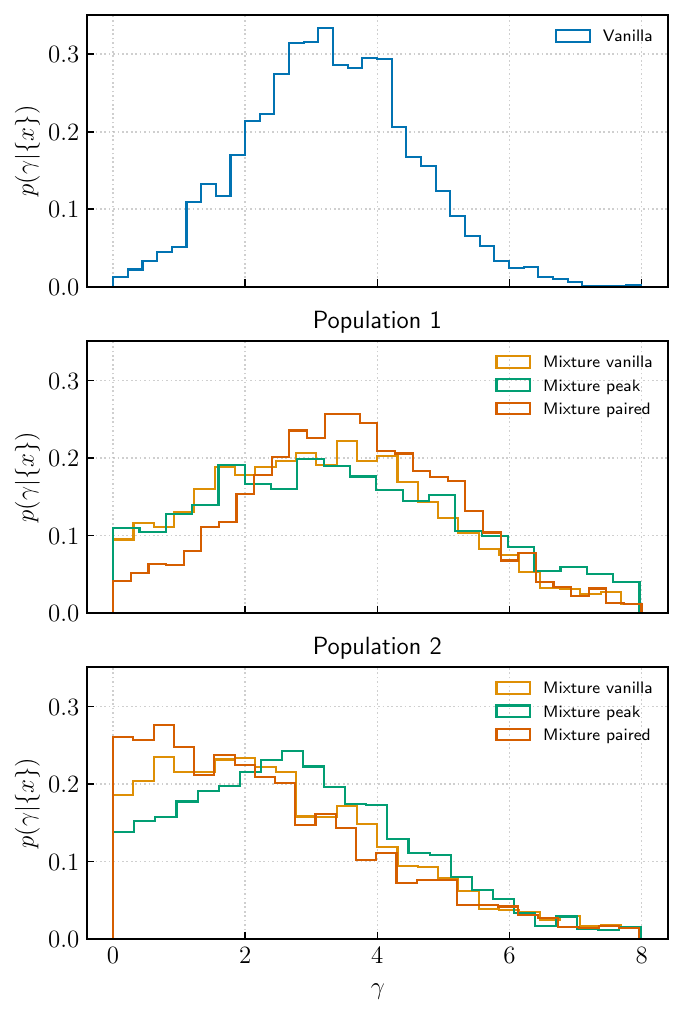}
    \caption{\textbf{CBC merger rate:} Histograms of the inferred posterior distribution of the CBC merger rate parameter $\gamma$, estimated for four distinct analysis on 59 real GW events from GWTC-2.1  GWTC-3 catalogs. Top: Vanilla analysis with a mono population model. Middle: Posterior of $\gamma$ for the first population of the \textsc{Mixture} analysis. Bottom: Posterior of $\gamma$ for the second population of the \textsc{Mixture} analysis.}
    \label{fig:gamma_mixturereal_ana}
\end{figure*}

Fig.~\ref{fig:mass_reconstructionreal_ana} shows the reconstructed mass spectrum obtained with the three \textsc{mixture} models; the inferred spectra are in perfect agreement with the canonical reconstruction of the non-evolving analysis, as well as with the mass spectrum reconstructed by \textsc{Evolving} models. Similarly, the two CBC merger rates inferred by the \textsc{mixture} model are all in agreement with each other and resemble the one of the baseline analysis. From Fig.~\ref{fig:gamma_mixturereal_ana}, we argue that the BBH merger rate as a function of redshift inferred for the first and second populations has probably the same trend. In other words, if the two sub-populations correspond to different formation channels, then both their time-delay distributions between the BBH formation and merger should be similar.

In order to validate the results observed with the \textsc{mixture} models, we ran the same population inference removing the spin distributions (masses and rate only). We find that the Bayes factor between the \textsc{mixture} model without spin and the simple non-evolving model is close to 1; this result highlights that the spin magnitude distribution is the determinant factor between the two sub-populations. Moreover, when removing the spin from the \textsc{mixture} population inference, the mixing parameter $\lambda_{\rm pop}$ is less constrained and can in agreement with 1, i.e., the presence of only one population.

\section{Sanity checks}
\label{sec:sanity}

As mentioned in the main text, to test and validate our new parametric models, as well as the population inference robustness, we performed two distinct: a simple mock-data-challenge (MDC) and a ``\textit{blurred}'' spin-mass analysis on real GW events.

\subsection{Simple Mock Data Challenge}

The aim of the MDC is to understand how Bayes factors and spin-mass models react to a population of BBHs that does not actually include any spin-mass relation. We simulated sets of $\sim 50$ detected GW events, in which we know that there is no spin-mass correlation. These events were drawn from the injection set used to estimate the selection effects in the main analysis and so they are representative of the sensitivity reached for GWTC--3. Here, we assume that we perfectly measure population parameters of the spins, masses, and distances in order to maximize the precision on the population parameters inferred by the different models. 

GW events are simulated following a MD-like BBH merger rate, a mass distribution following a \textsc{Power Law + peak} vanilla model and a spin distribution preferring lowly spinning BBHs nearly aligned with the orbital angular momentum. The distributions of masses and spins are indicated with a black dashed line in the figures referenced below.

We report a summary of the results under the form of Bayes factors in the upper half of Tab.~\ref{Tab:Bayes factor MDC SWAP}.

\begin{table}[h]
    \centering
    \begin{tabular}{l c c} 
    \textbf{Mock Data Challenge} & &\\ \\
    \hline
    \textbf{Model} & $\boldsymbol{\log_{10}\mathcal{B}}$ & \hspace{0.5cm}$\boldsymbol{\log_{10}\mathcal{L}_{\rm max}}$ \\ [0.9ex] 
    \hline
    \textsc{Mixture vanilla}  & -1.36 & 2.53 \\ 
    \textsc{Mixture peak} & -3.69 & -0.88 \\ 
    \textsc{Mixture paired} & -3.26 & 3.16 \\ 
    \textsc{Evolving Gaussian}  & -6.07 & -1.46 \\ 
    \textsc{Beta to Gaussian} & -0.61 & 1.86 \\ 
    \textsc{Beta to Beta}  & -0.21 & 0.43 \\
    \hline \hline \\
    \textbf{Spin-mass blurred analysis} &  &\\ 
    \textbf{(real data)} &  &\\ \\
    \hline
    \textbf{Model} & $\boldsymbol{\log_{10}\mathcal{B}}$ & \hspace{0.5cm}$\boldsymbol{\log_{10}\mathcal{L}_{\rm max}}$ \\ [0.9ex] 
    \hline
    
    \textsc{Mixture vanilla}  & 0.90 & 0.92 \\ 
    \textsc{Mixture peak}  & 0.76 & 0.40 \\ 
    \textsc{Mixture paired}  & 1.11 & 2.24 \\ 
    \textsc{Evolving Gaussian}  & -3.34 & -0.23 \\ 
    \textsc{Beta to Gaussian}  & -0.28 & 0.18 \\ 
    \textsc{Beta to Beta} & 0.06 & 0.40 \\ 
     
    \hline \hline   
    \end{tabular}
    \caption{\textbf{Model assessment:} Log-Bayes factors (second column) and the maximum of log-likelihood ratio (third column) obtained for each of our six models, compared to the canonical non evolving model when inferring the population parameters of the mock data challenge (MDC) and the swapping test (SWAP)}
    \label{Tab:Bayes factor MDC SWAP}
\end{table}
From the Bayes factor obtained for the MDC analysis, it is clear that the canonical non-evolving model is always preferred for the \textsc{mixture} and \textsc{evolving gaussian} models and equally preferred for the \textsc{Beta to Beta} and \textsc{Beta to gaussian} models. This result demonstrates that the spin-mass correlation directly drives the model selection in the real analysis presented in the real set of GW data. 

\subsubsection{Considerations on Evolving models} 
\begin{figure*}
    \centering
    \includegraphics[width=\textwidth]{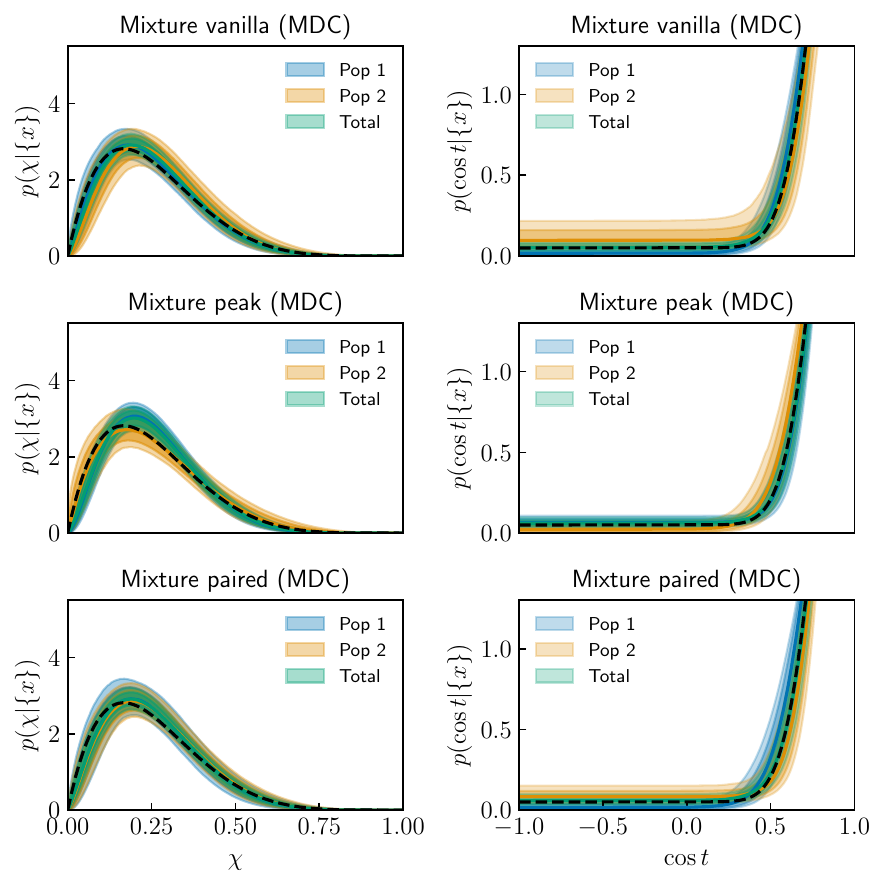}
    \caption{\textbf{Spin population spectra (MDC):} Reconstructed spectra of the spin magnitude $\chi$ (left) and the cosine of the tilt angle (right), obtained from the population inference of simulated GW data with no spin-mass evolution (MDC) using the \textsc{Mixture} models. The plain lines are the median of the reconstructed spectra and the colored contours are the $90\%$ C.L. And the dotted black line is the true distribution in the data.}
    \label{fig:mixture spin mag MDC}
\end{figure*}

\begin{figure*}
    \centering
    \includegraphics[width=0.7\textwidth]{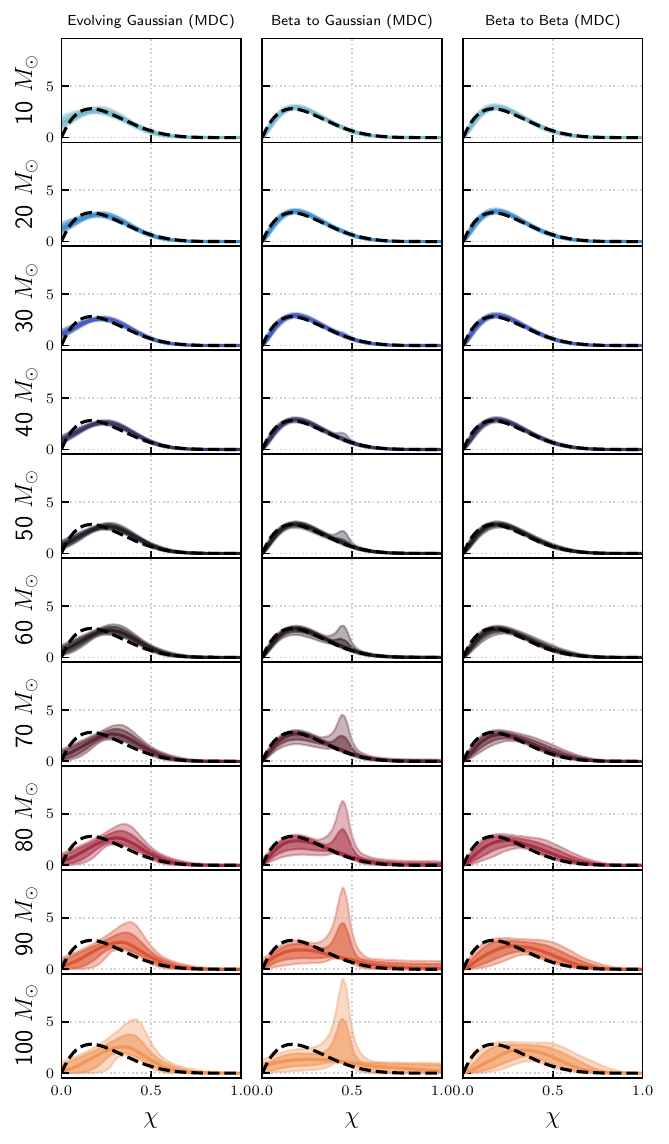}
    \caption{\textbf{Evolving spin magnitude spectra (MDC):} Joy plot of the reconstructed spin magnitude spectra obtained with the \textsc{Evolving gaussian} (left) and \textsc{Evolving} window (middle and right) models on simulated GW data (MDC). The black line is the true distribution in the data. The colored contours are the reconstructed spectra defined as the $90\%$ C.L. The discrepancies in the higher mass regime are induced by the prior effects since the inferred transition mass parameter posterior $\rm m_{t}$ is railing on its prior range. This effect is not to worry in the main analysis since the posterior nicely converges in its prior range.}
    \label{fig:evolving spin mag MDC}
\end{figure*}

Fig.~\ref{fig:mixture spin mag MDC} and Fig.~\ref{fig:evolving spin mag MDC} are the reconstructed spin magnitudes and tilt angles obtained from the \textsc{mixture} and \textsc{evolving} models. For the \textsc{mixture} model, the results demonstrate strong agreement with the original data, accurately capturing the true spin distributions. Similarly, the \textsc{evolving} model also yields excellent results, correctly identifying the injected population's spin magnitudes. The deviations in the posterior predictive checks above 70 $M_\odot$ are due to the prior ranges on $m_{\rm t}$ and $\delta m_{\rm t}$ governing the window function. Specifically, as no transition mass is present for the spin population, $m_{\rm t}$ is not bounded within the upper limit of its prior range (100 $M_\odot$). As a consequence, the posterior predictive check of the window function, see Fig.~\ref{fig:window MDC and Swap}, still has some support for a transition happening below 100 $M_\odot$.

\begin{figure*}
    \centering
    \includegraphics[width=0.7\textwidth]{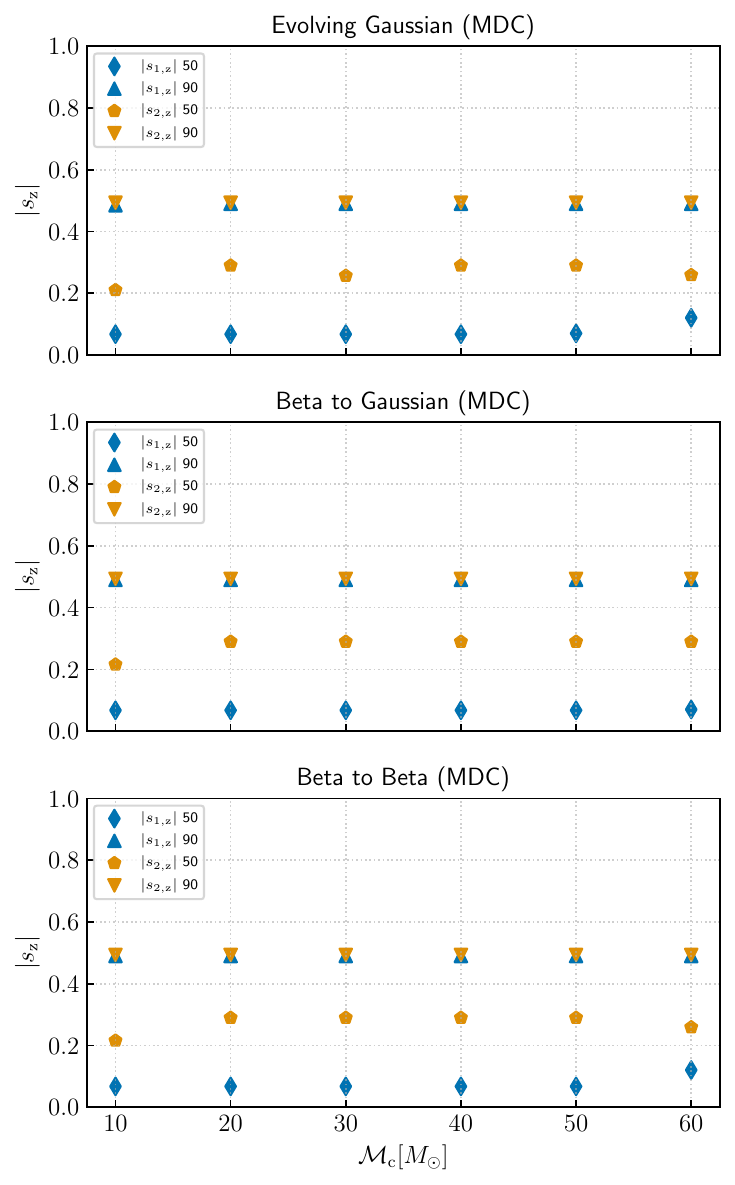}
    \caption{\textbf{Aligned spin magnitude evolution MDC:} Scatter plot representing the evolution of the aligned component of the spin magnitude $\rm s_{z}$ with respect to the chirp mass $\mathcal{M}_c$, obtained from the population inference of simulated GW data (MDC) using the \textsc{Evolving Gaussian} (top), the \textsc{Beta to Gaussian} (middle) and \textsc{Beta to Beta} (bottom) models.}
    \label{fig:chirp mass spin MDC}
\end{figure*}

In analogy to the real analysis, we also replicate the plot on the $|s_z|$ detector chirp mass relation. Fig. \ref{fig:chirp mass spin MDC} shows us that when no spin-mass relation is present, and GWs masses and spins are well-measured, no trend can be observed between these two variables. 

\begin{figure*}
    \centering
    \includegraphics[width=\textwidth]{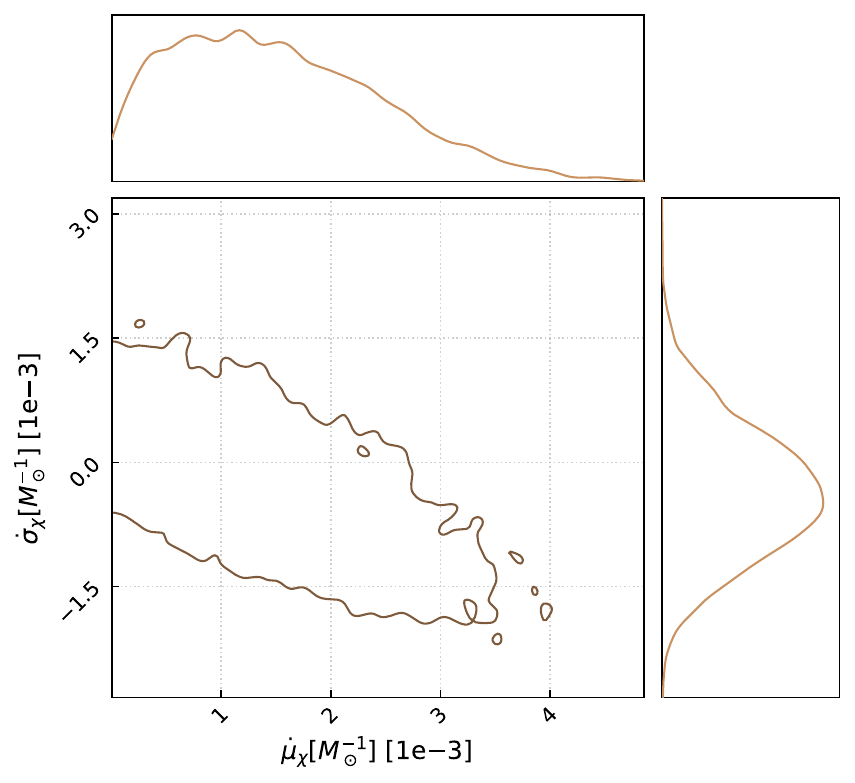}
    \caption{\textbf{Evolving Gaussian parameter MDC:} Corner plot of the population parameters governing the evolution of the mean and width of the Gaussian distribution modelling the spin magnitude in the \textsc{Evolving Gaussian} model, namely $\dot{\mu}_{\chi}$ and $\dot{\sigma}_{\chi}$.}
    \label{fig:mu dot sigma dot MDC}
\end{figure*}

Concerning the \textsc{Evolving gaussian} model, Fig.~\ref{fig:mu dot sigma dot MDC} shows the inferred distribution for the population parameters driving evolution of the spin mean and with. We can observe that, contrary to the real analysis, the model correctly supports a non-evolution of the spin magnitude with the mass.

\subsubsection{Considerations on mixing models}

\begin{figure*}
    \centering
    \includegraphics[width=0.6\textwidth]{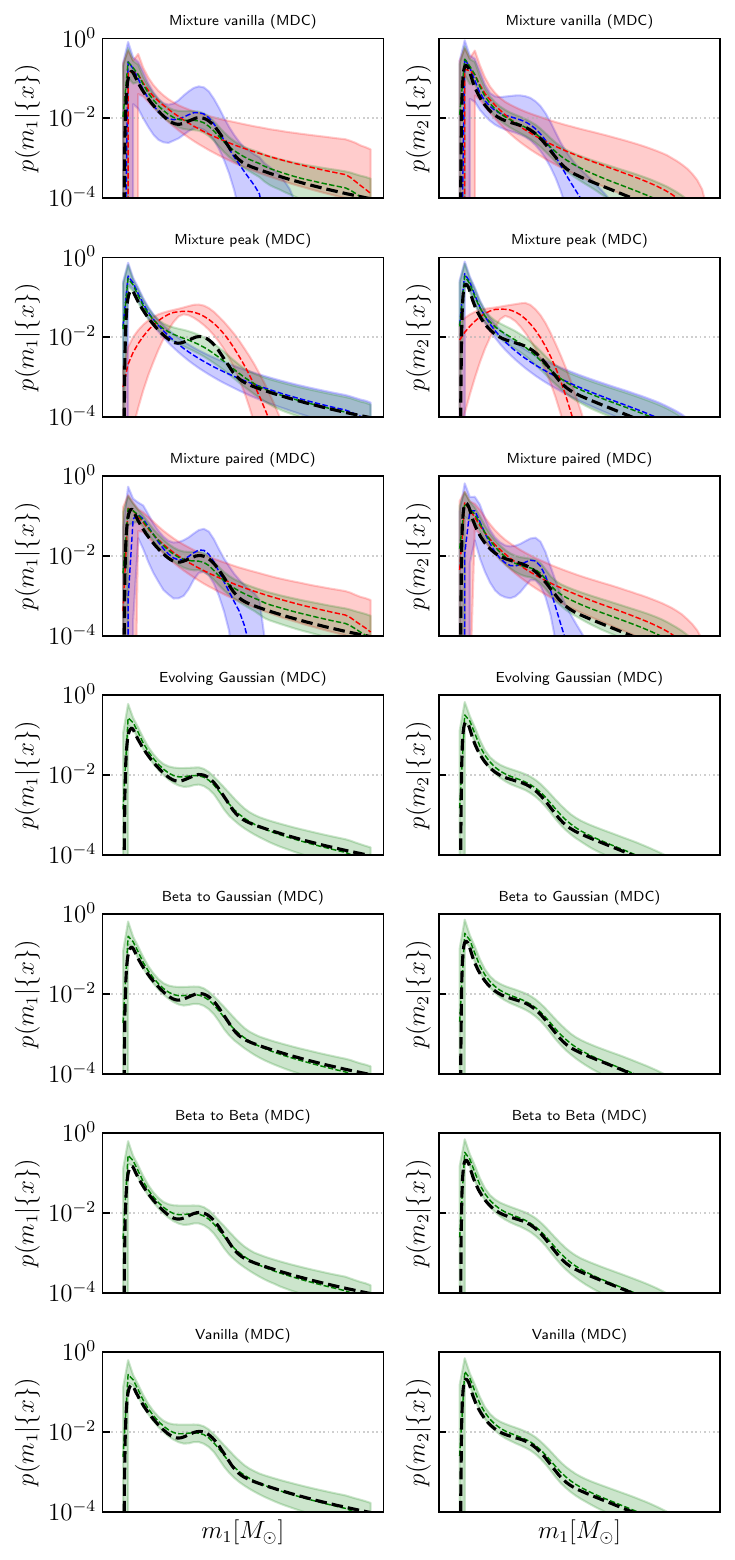}
    \caption{\textbf{Mass spectrum reconstruction MDC:} Reconstructed mass spectra of the primary (left column) and secondary (right column) masses obtained from MDC population inference of the MDC  with the \textsc{Mixture vanilla} model (first row), \textsc{Mixture peak} (second row), \textsc{Mixture paired} (third row), \textsc{Evolving gaussian} (fourth row), \textsc{Beta to Gaussian} (fifth row), \textsc{Beta to Beta} (sixth row) and the canonical \textsc{Vanilla} model (seventh row). The dotted lines are the median values and the colored contours the $90\%$ C.L. inferred.}
    \label{fig:mass spectrum MDC}
\end{figure*}

For \textsc{mixture} models, in Figs. \ref{fig:mixing param MDC and swap}-\ref{fig:bi modality MDC and swap}, we studied the response of $\lambda_{\rm pop}$ and its interplay with the minimum mass of the secondary sub-population. For the \textsc{Mixture vanilla} and \textsc{Mixture paired} analysis, the inference is not able to distinguish the presence of the two populations as the posteriors on the mixture fraction are informative. The minimum mass of the secondary population also strongly supports a value corresponding to the overall minimum of the true population. For the \textsc{Mixture peak model}, the mixture fraction is constrained as the gaussian peak is used to fit the central peak of the \textsc{Power Law + peak} distribution that has been simulated. This can be clearly seen in Fig. \ref{fig:mass spectrum MDC} where we report the mass posterior predictive checks from the models in overlap with the simulated distribution. The posteriors on the CBC merger rate parameter $\gamma$ are shown in Fig. \ref{fig:merger rate MDC and swap} and they are all consistent with the simulated value of 2.7.

\subsection{Spin-mass blurred analysis with real data}

Another sanity check that we perform is to repeat all of our analyses on a spin-mass blinded data set. We select the same GW events as the ones considered for the real analysis, but this time we permute among them their inferred spin values. In the limit that the determination of the other GW parameters, and the selection biases, are not strongly related to the spin magnitudes, this procedure artificially blinds the dataset to any spin-mass relation. 

\begin{figure*}
    \centering
    \includegraphics[width=\textwidth]{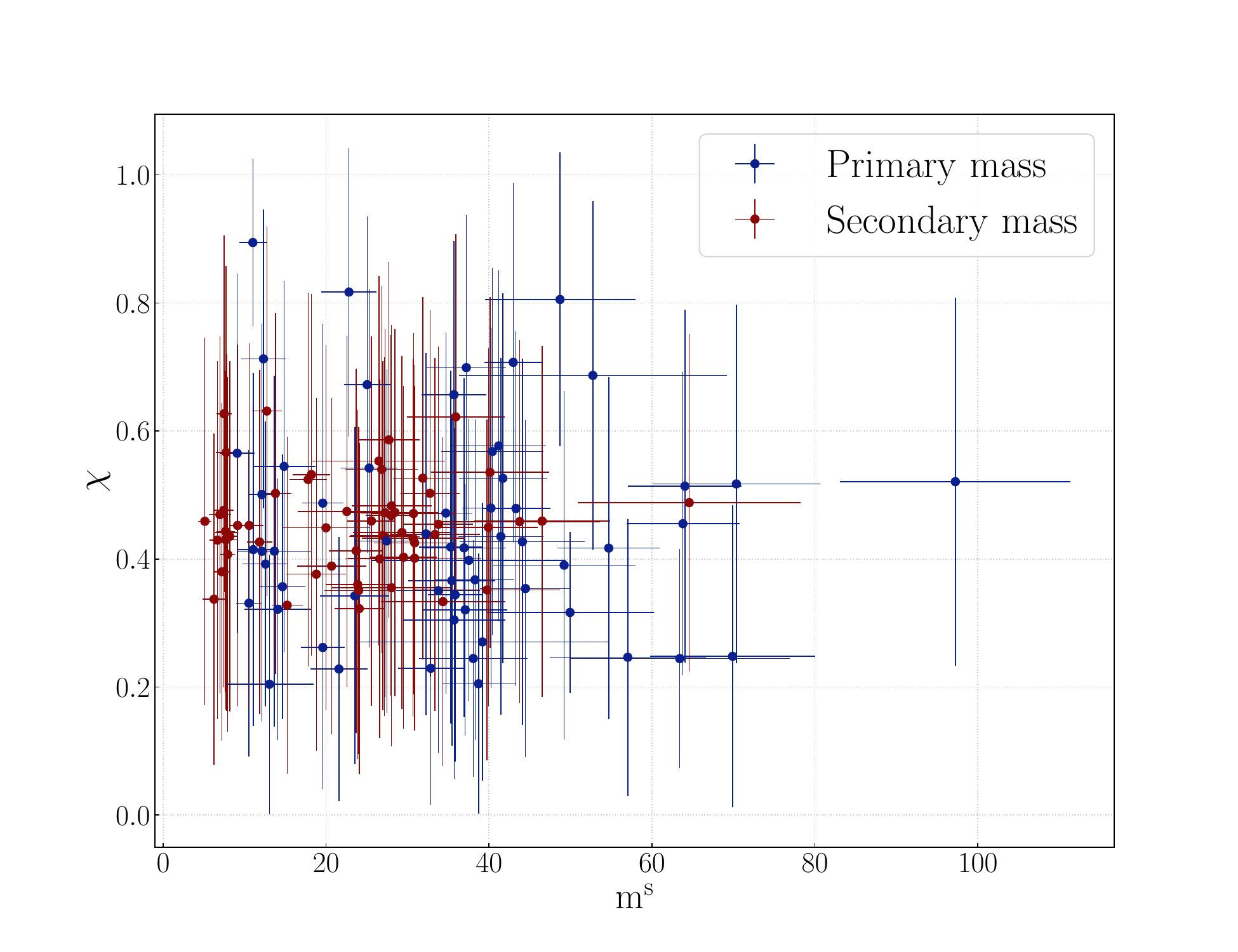}
    \caption{\textbf{Blurred Mass-Spin dataset:} Scatter plot of the GW events used in the analysis. They correspond to the 59 binary black hole events from the GWTC-2.1  GWTC-3 catalogs selected with an $\rm IFAR\geq1 yr$. The x-axis shows the source frame masses $\rm m^{\rm s}$ and the y-axis displays the dimensionless spin magnitudes $\chi$. The errors bars are the $1\sigma$ uncertainties of the official LVK parameter estimation samples for each GW event.}
    \label{fig:swap scatter plot mass source and chi}
\end{figure*}

We warrant that in the low number detections regime, if there is really a spin-mass relation, shuffling the inferred spin values might not 100\% blind the data from the spin-mass relation. The motivation is that low-mass events are more numerous, therefore there is an higher probability that the spin of a low-mass event will be reassigned to a low-mass event thus preserving the spin-mass correlation. However, the shuffling of the spin values is expected for sure to blind any spin-mass relation at high masses (since the events are less numerous). We refer to this type of analysis as spin-mass ``blurred'' analysis to indicate that we are not able to 100\% generate blinded data. The spin-mass values used for this analysis are displayed in Fig.~\ref{fig:swap scatter plot mass source and chi} for the blurred spin-mass estimations.

The Bayes factors and log-likelihood ratios obtained from the blinded analyses are reported in Tab.~\ref{Tab:Bayes factor MDC SWAP}. All the models report inconclusive results, having no preference for any of the evolving spin-mass models. In terms of population constraints for the different models, we find very similar results as the ones described for the MDC. 

\subsubsection{Considerations on Evolving models}

\begin{figure*}
    \centering
    \includegraphics[width=0.7\textwidth]{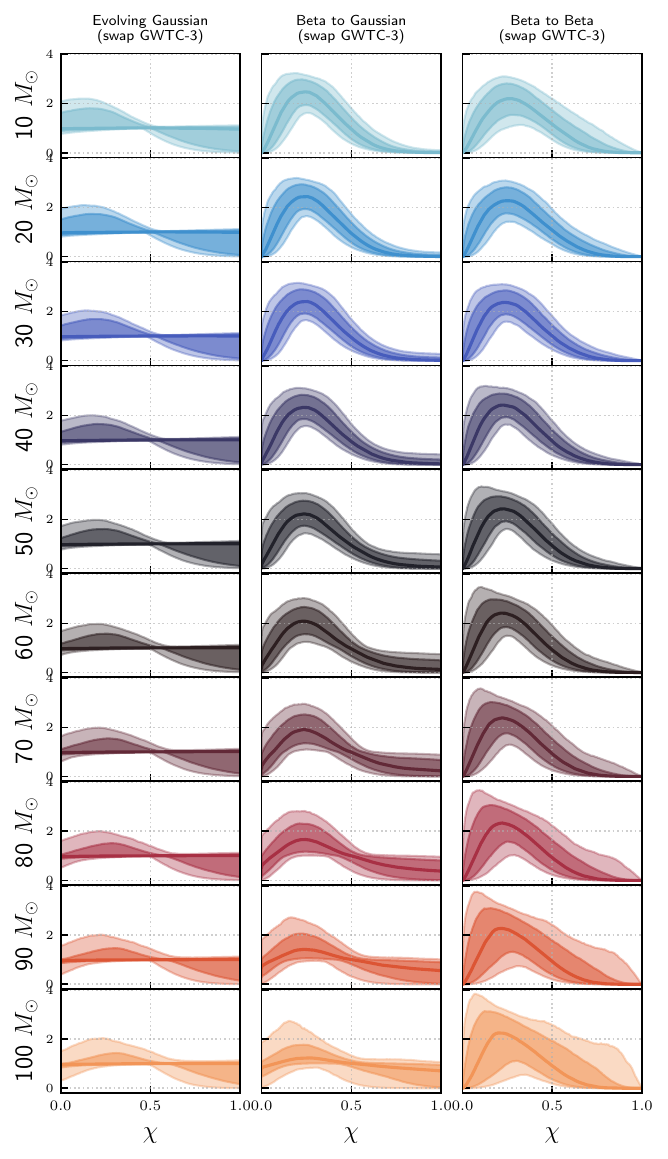}
    \caption{\textbf{Mixture spin magnitude spectra (blurred):} Joy plot of the reconstructed spin magnitude spectra obtained with the \textsc{Evolving gaussian} (left) and \textsc{Evolving} window models (middle and right) on 59 real GW events from GWTC-2.1  GWTC-3 catalogs, shuffled to remove the spin-mass correlation (blurred). The colored contours are the the $90\%$ C.L. reconstructed spectra from the population inference.}
    \label{fig:spin mag evolving swap}
\end{figure*}

Fig.~\ref{fig:spin mag evolving swap} shows the posterior predictive check for the three \textsc{evolving} models. In comparison with the real analysis, we still observe that at low masses the analysis reconstructs a spin distribution preferring low spin values, but the reconstruction becomes more and more uncertain as the mass value increases. This is to the fact that as low-mass events are more numerous, the spin shuffling does not effectively blind the data to a spin-mass relation. However, the spin-mass relation is diluted enough for the Bayes factors on model selection to become inconclusive.

\begin{figure*}
    \centering
    \includegraphics[width=\textwidth]{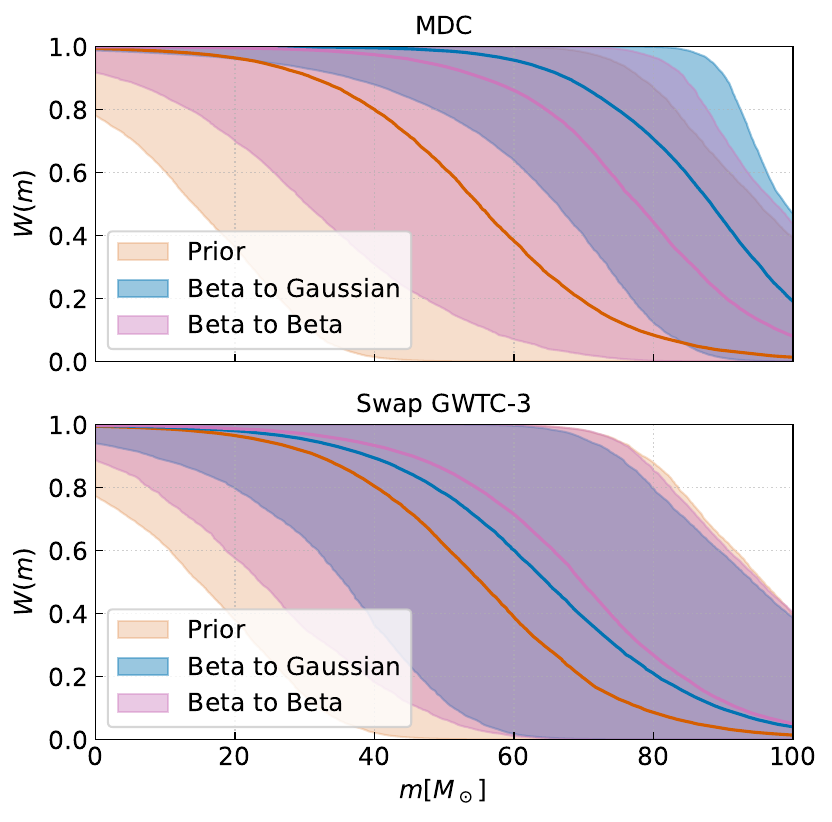}
    \caption{\textbf{Window function (MDC and blurred):} Reconstructed window function for the MDC (top) and the blurred (bottom) sanity checks, with the \textsc{Beta to Gaussian} and \textsc{Beta to Beta}. The colored contours are the $90\%$ C.L inferred spectra.}
    \label{fig:window MDC and Swap}
\end{figure*}

\begin{figure*}
    \centering
    \includegraphics[width=\textwidth]{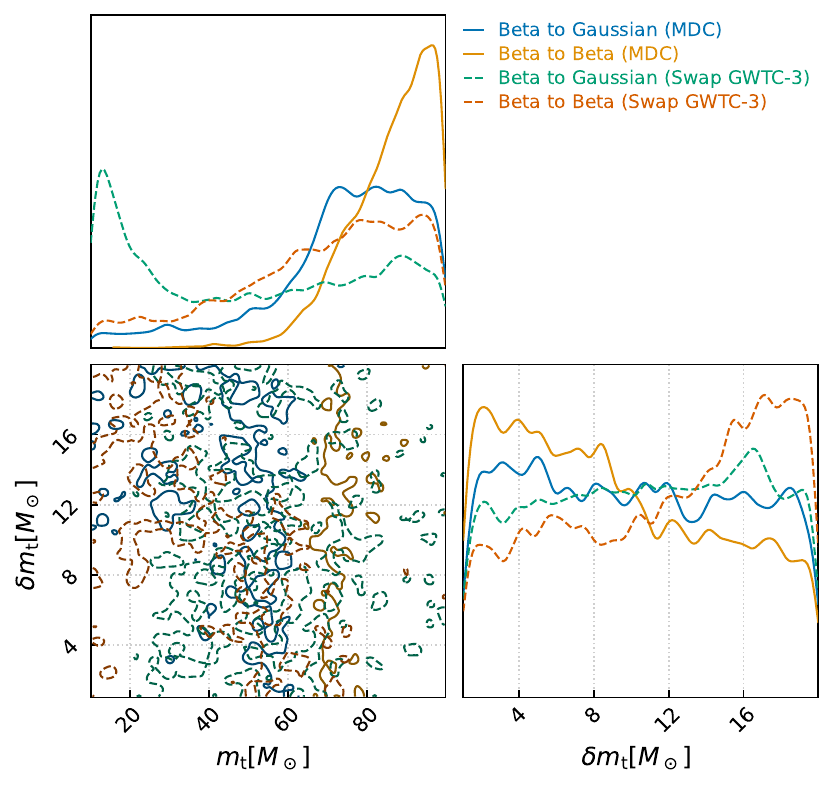}
    \caption{\textbf{Spin-mass transition (MDC and blurred):} Overlapped corner plots of the population parameters governing the spin transition as a function of the mass, namely $\rm m_{t}$ and $\delta_{\rm m_{t}}$; for the two \textsc{Shifting} models in the case of the MDC and the blurred analysis.}
    \label{fig:mt and delta mt MDC and swap}
\end{figure*}

\begin{figure*}
    \centering
    \includegraphics[width=0.7\textwidth]{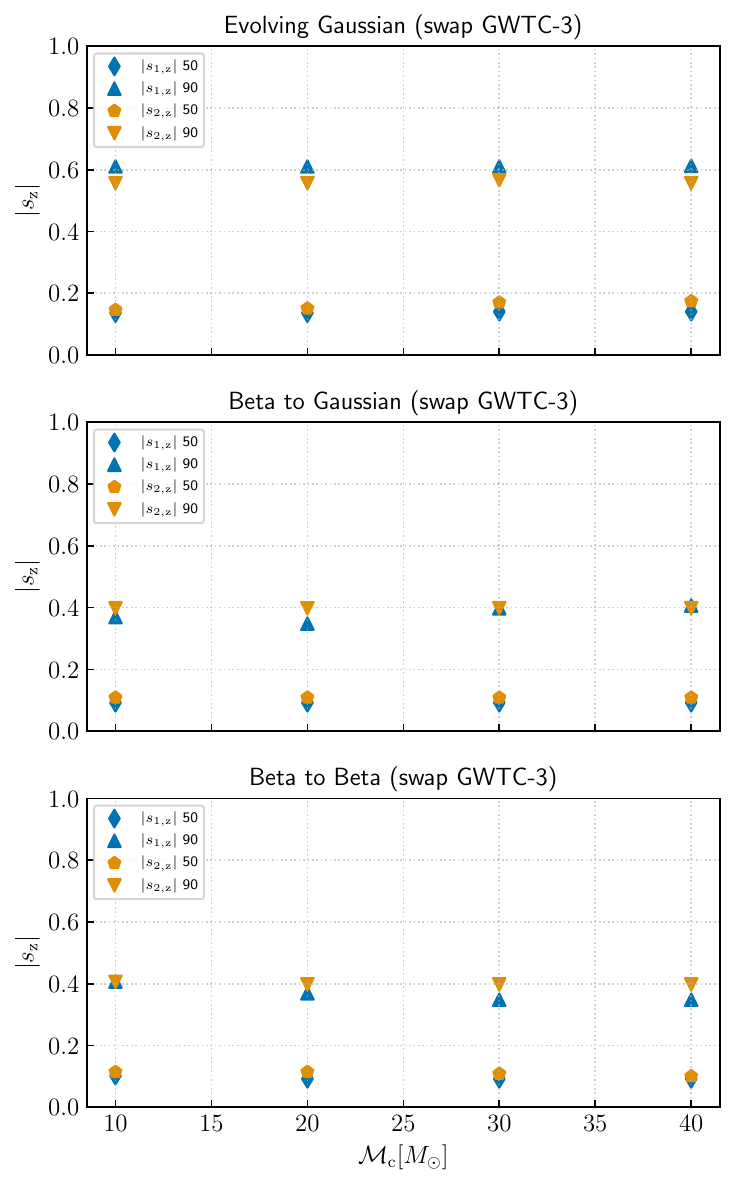}
    \caption{\textbf{Aligned spin magnitude evolution blurred:} Scatter plot representing the evolution of the aligned component of the spin magnitude $\rm s_{z}$ with respect to the chirp mass $\mathcal{M}_c$, obtained from the population inference of real but shuffled GW data (swap) using the \textsc{Evolving Gaussian} (top), the \textsc{Beta to Gaussian} (middle) and \textsc{Beta to Beta} (bottom) models.}
    \label{fig:chirp mass spin swap}
\end{figure*}

Concerning the \textsc{beta to gaussian} and \textsc{beta to beta} model, Fig.~\ref{fig:window MDC and Swap} (bottom panel) and Fig.~\ref{fig:mt and delta mt MDC and swap} show the posterior predictive check and population posteriors on the window function. For the real analysis, we observe that the window function supports a transition at higher masses, which is a hint of the fact that the spin shuffling procedure has partially hidden the spin-mass relation in real data. Also for this test case, in Fig.~\ref{fig:chirp mass spin swap} we study the trend between $|s_{z}|$ and the detector chirp mass, finding no evident relation as the one observed in the real analysis.

\begin{figure*}
    \centering
    \includegraphics[width=\textwidth]{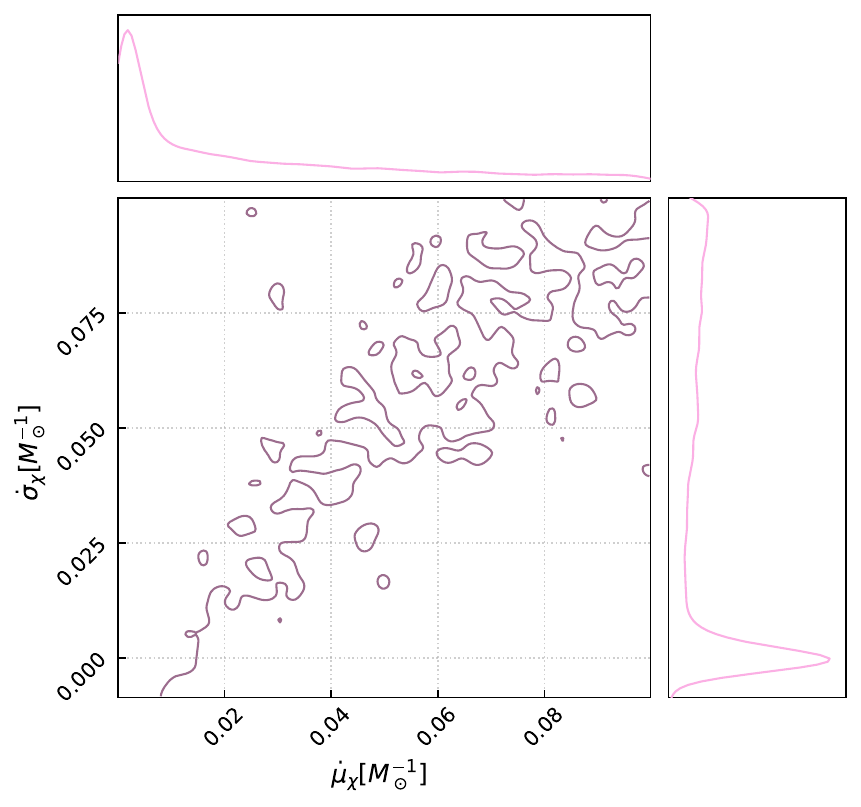}
    \caption{\textbf{Evolving Gaussian parameter blurred:} Corner plot of the population parameters governing the evolution of the mean and width of the Gaussian distribution modelling the spin magnitude in the \textsc{Evolving Gaussian} model, namely $\dot{\mu}_{\chi}$ and $\dot{\sigma}_{\chi}$.}
    \label{fig:mu dot sigma dot swap}
\end{figure*}

Regarding the \textsc{evolving gaussian} model, we find that there is no support for a continuous evolution of the distribution of the spin magnitude. Fig.~\ref{fig:mu dot sigma dot swap} shows the posterior on parameters governing the evolution of the mean and standard deviation of the gaussian spin distribution. 

\subsubsection{Considerations on the mixture models}

\begin{figure*}
    \centering
    \includegraphics[width=\textwidth]{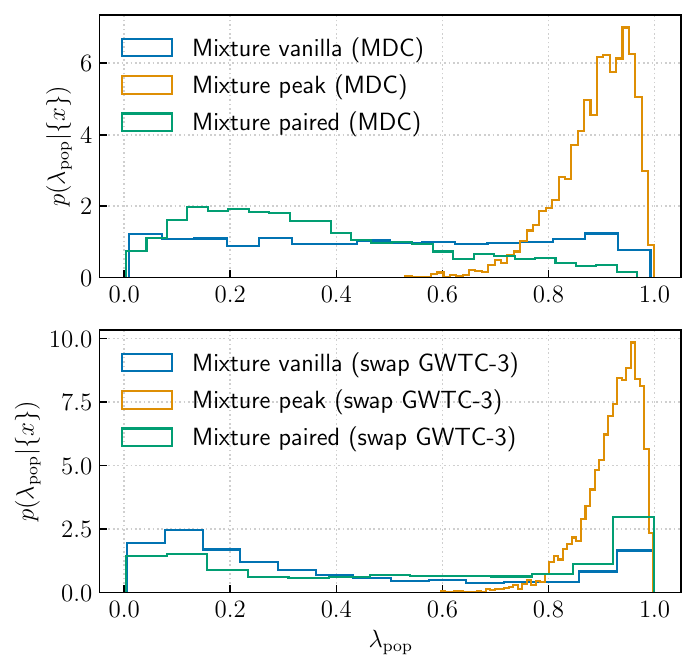}
    \caption{\textbf{Mixing parameter (MDC and blurred):} (Top plot) posteriors distribution of the inferred $\lambda_{\rm pop}$ from the MDC analysis using the three flavours of the \textsc{Mixture} model. (Bottom plot) posteriors distribution of the inferred $\lambda_{\rm pop}$ from the blurred analysis using the three flavours of the \textsc{Mixture} model. }
    \label{fig:mixing param MDC and swap}
\end{figure*}

\begin{figure*}
    \centering
    \includegraphics[width=\textwidth]{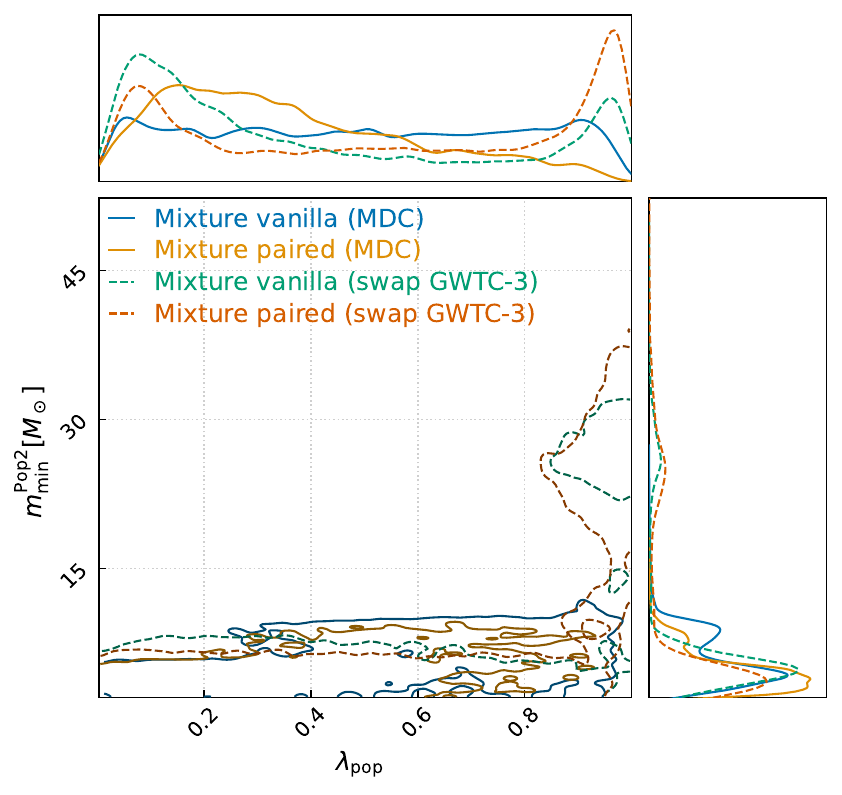}
    \caption{\textbf{Mixture-mass relation:} Overlapped corner plots of the mixing parameter $\lambda_{\rm pop}$ and the minimum mass of the secondary population $\rm m_{\rm min}^{\rm pop2}$, obtained from the \textsc{Mixture vanilla} and \textsc{Mixture paired} models from the MDC and the blurred analysis.}
    \label{fig:bi modality MDC and swap}
\end{figure*}

\begin{figure*}
    \centering
    \includegraphics[width=\textwidth]{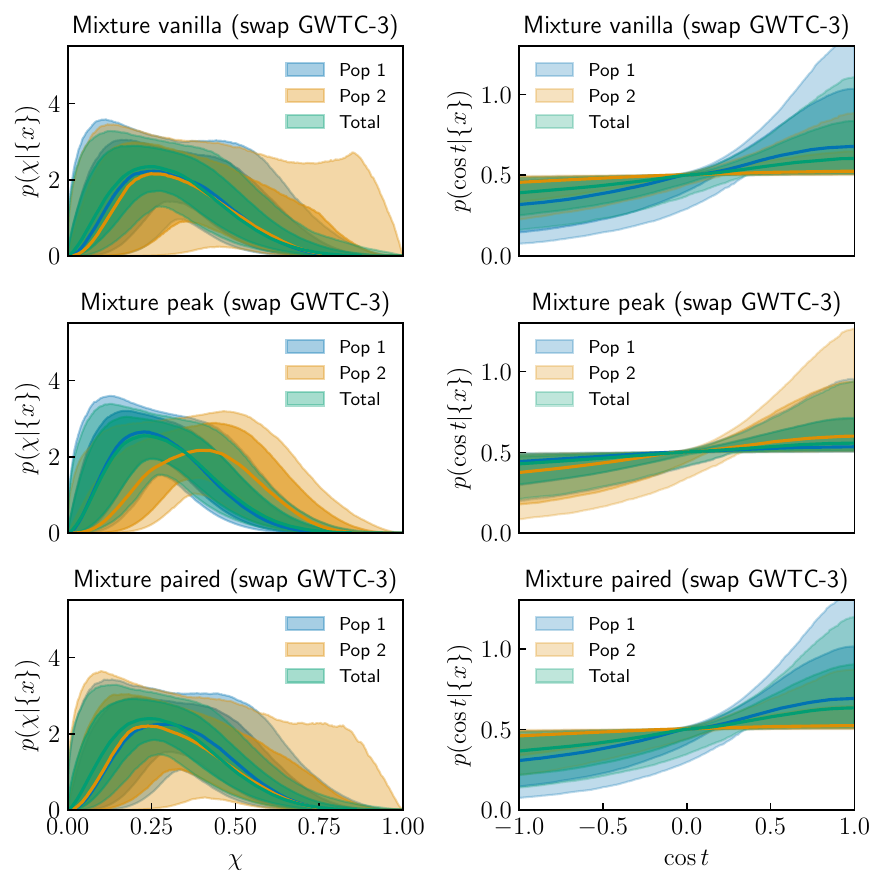}
    \caption{\textbf{Spin population spectra (blurred):} Reconstructed spectra of the spin magnitude $\chi$ (left) and the cosine of the tilt angle (right), obtained from the population inference of 59 real GW events from GWTC-2.1  GWTC-3 catalogs which as been shuffled to remove any spin-mass correlation (blurred). The plain lines are the median of the reconstructed spectra and the colored contours are the $90\%$ C.L. And the dotted black line is the true distribution in the data.}
    \label{fig:spin mag mixture swap}
\end{figure*}

In Fig.~\ref{fig:spin mag mixture swap} we show the posterior predictive checks for the spin magnitude and tilt angle distributions. In this case, both sub-populations support low spin values with the secondary population (more massive events) slightly more uncertain. Similar to the MDC test, the mixing parameter $\lambda_{\rm pop}$ shown in Fig.~\ref{fig:mixing param MDC and swap} and Fig.~\ref{fig:bi modality MDC and swap} is not able to pinpoint the presence of any sub-population as opposed to what we observed with unblinded GW data.

\begin{figure*}
    \centering
    \includegraphics[width=0.85\textwidth]{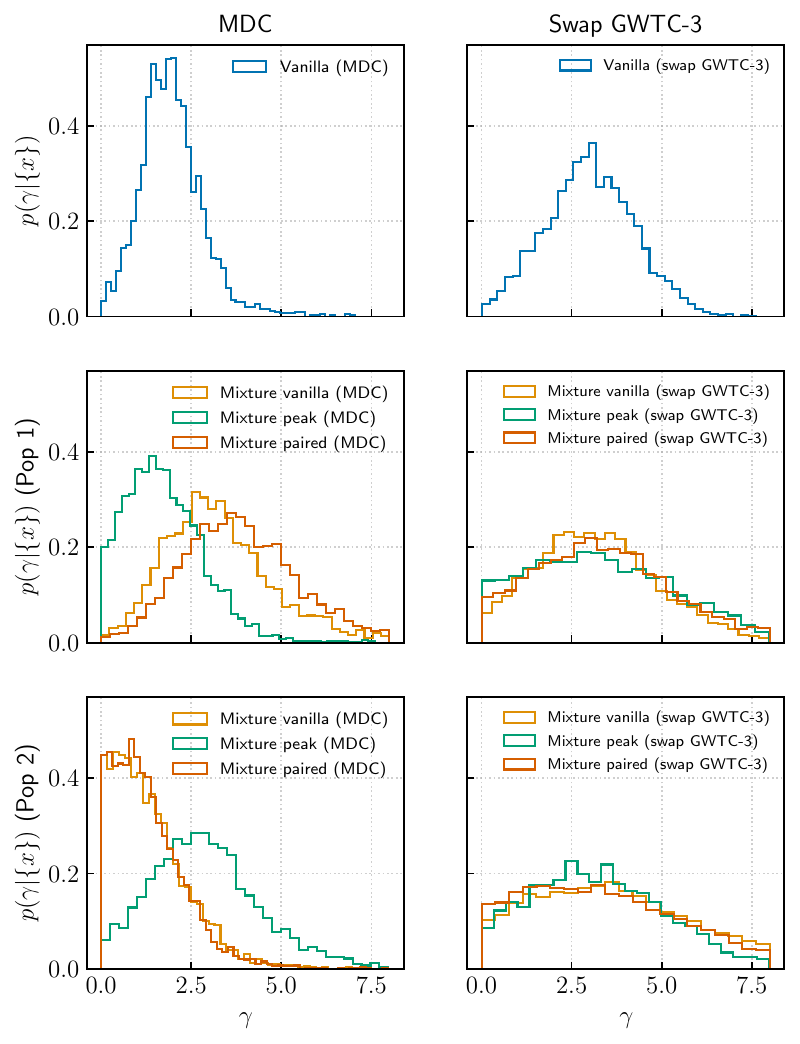}
    \caption{\textbf{CBC merger rate (MDC and blurred):} Inferred posterior of the $\gamma$ parameter from the CBC merger rate population model. Left plot: Estimated $\gamma$ from the MDC with the non evolving canonical model (top), first population with \textsc{Mixture} models (middle) and second population with the \textsc{Mixture} models (bottom). Right plot: Estimated $\gamma$ from the blurred analysis with the non evolving canonical model (top), first population with \textsc{Mixture} models (middle) and second population with the \textsc{Mixture} models (bottom) }
    \label{fig:merger rate MDC and swap}
\end{figure*}

\begin{figure*}
    \centering
    \includegraphics[width=0.6\textwidth]{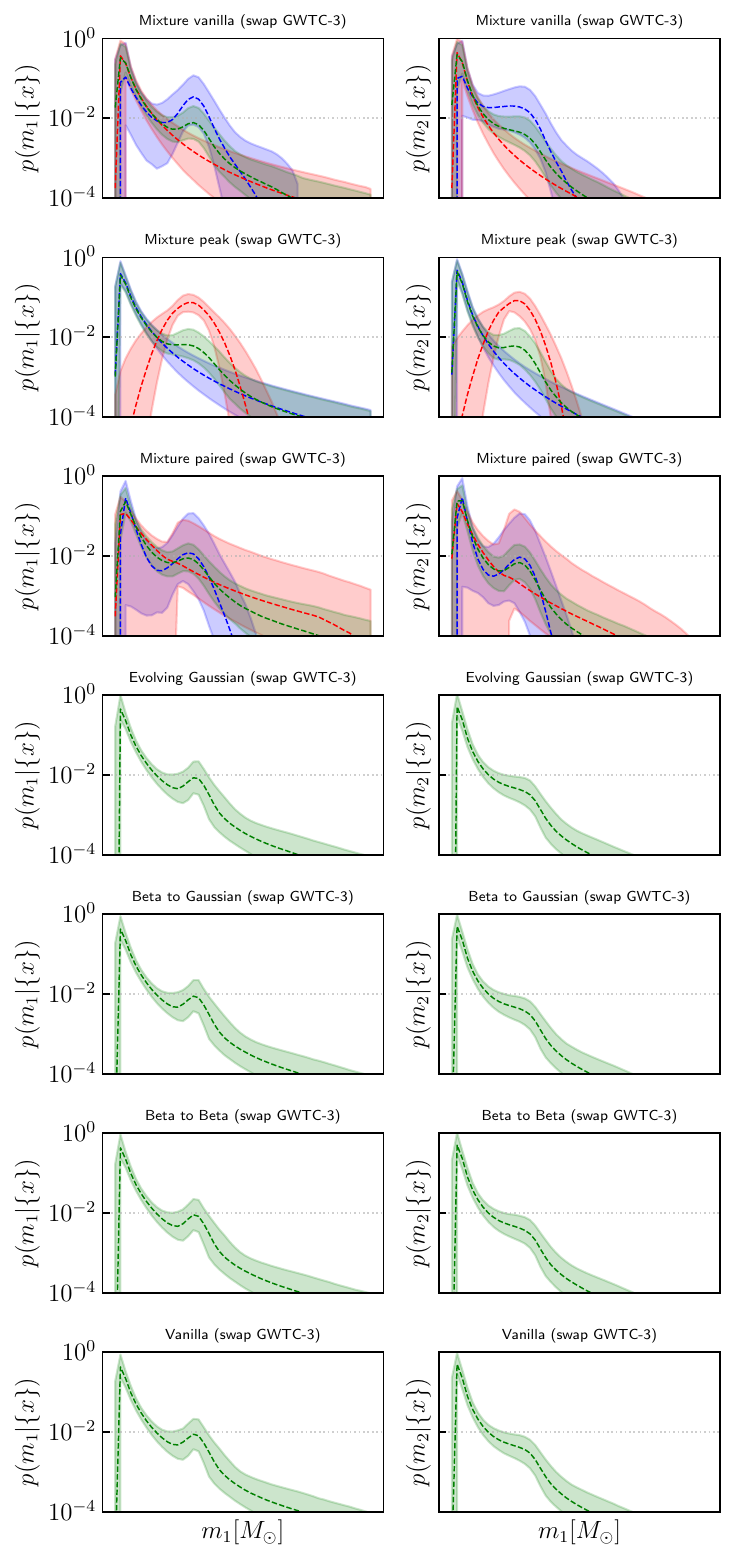}
    \caption{\textbf{Mass spectrum reconstruction blurred:} Reconstructed mass spectra of the primary (left column) and secondary (right column) masses obtained from blurred population inference of the MDC  with the \textsc{Mixture vanilla} model (first row), \textsc{Mixture peak} (second row), \textsc{Mixture paired} (third row), \textsc{Evolving gaussian} (fourth row), \textsc{Beta to Gaussian} (fifth row), \textsc{Beta to Beta} (sixth row) and the canonical \textsc{Vanilla} model (seventh row). The dotted lines are the median values and the colored contours the $90\%$ C.L. inferred.}
    \label{fig:mass spectrum swap}
\end{figure*}

The mass spectrum reconstructed in this test still agrees with the one reconstructed in the real analysis thus indicating that the spin-mass relation is not an artifact created by the fit of the mass spectrum. Figure~\ref{fig:mass spectrum swap} presents the reconstructed mass spectrum for both model families, showing good agreement among spectra and with the non-evolving analysis. This agreement extends to the inferred compact binary coalescence (CBC) merger rate, as depicted in Fig. \ref{fig:merger rate MDC and swap}.

In conclusion, when analyzing real GW data devoid of spin-mass correlation, the mixture and evolving models successfully reconstruct the correct mass, merger rate, and spin spectra, while indicating no support for spin evolution. Additionally, these models are disfavored by the Bayes factors.

\subsection{Numerical stability tests}
\begin{figure*}
    \centering
    \includegraphics[width=1.\textwidth]{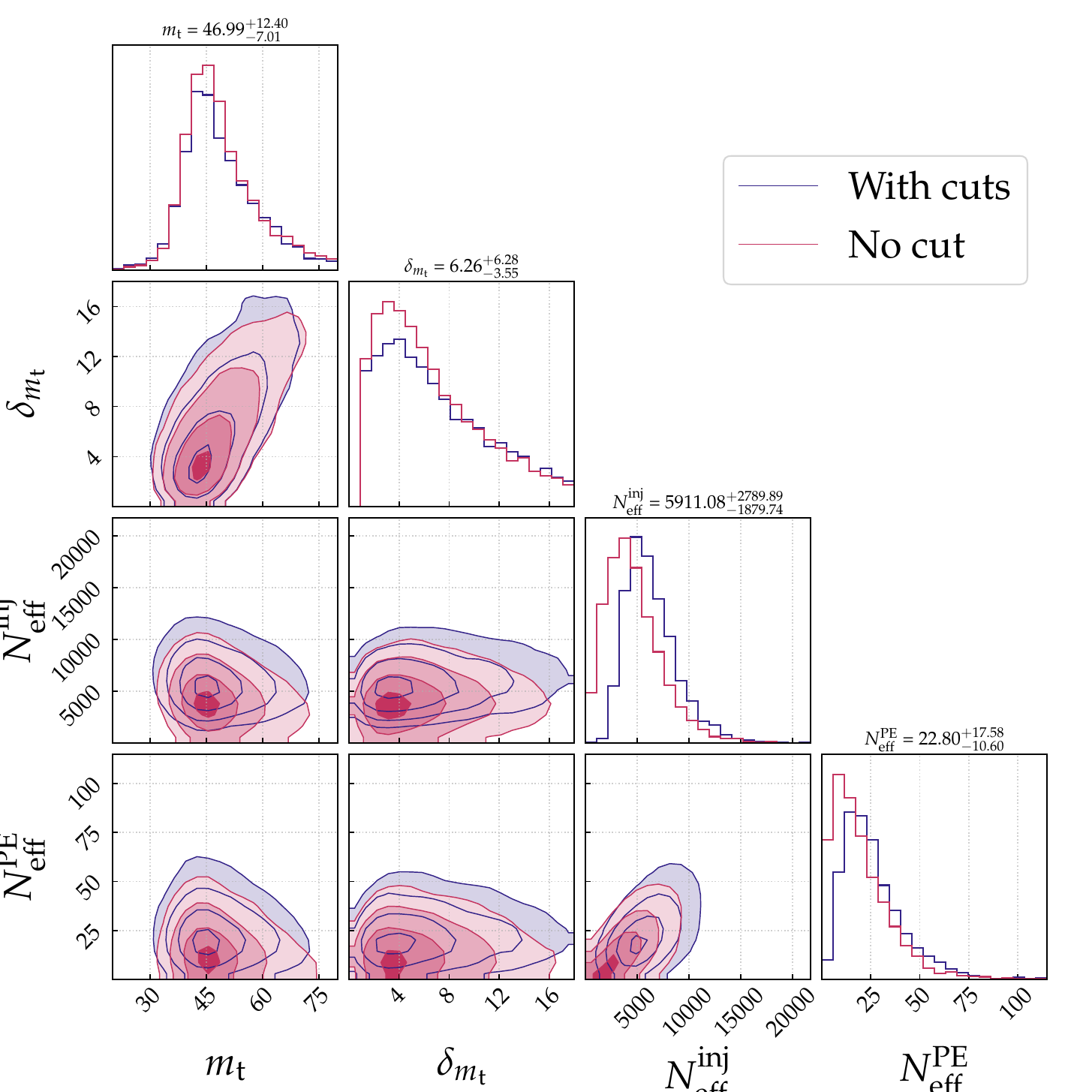}
    \caption{\textbf{Corner plot \textsc{Evolving model}:} Corner plot of the population parameters ($m_{\rm t}$,$\delta_{\rm m_{t}}$) and the stability estimator ($N_{\rm eff}^{\rm inj}$,$N_{\rm eff}^{\rm PE}$), obtained with the \textsc{Beta to Beta} evolving model on the 59 GW events with $\rm IFAR \geq 1 yr$. The purple contours and histograms were estimated while putting the minimum value of $N_{\rm eff}^{\rm inj}=200$ and $N_{\rm eff}^{\rm PE}=10$. The pink contours and histograms were estimated while placing not cuts on these estimators.}
    \label{fig:corner BtB}
\end{figure*}

\begin{figure*}
    \centering
    \includegraphics[width=1.\textwidth]{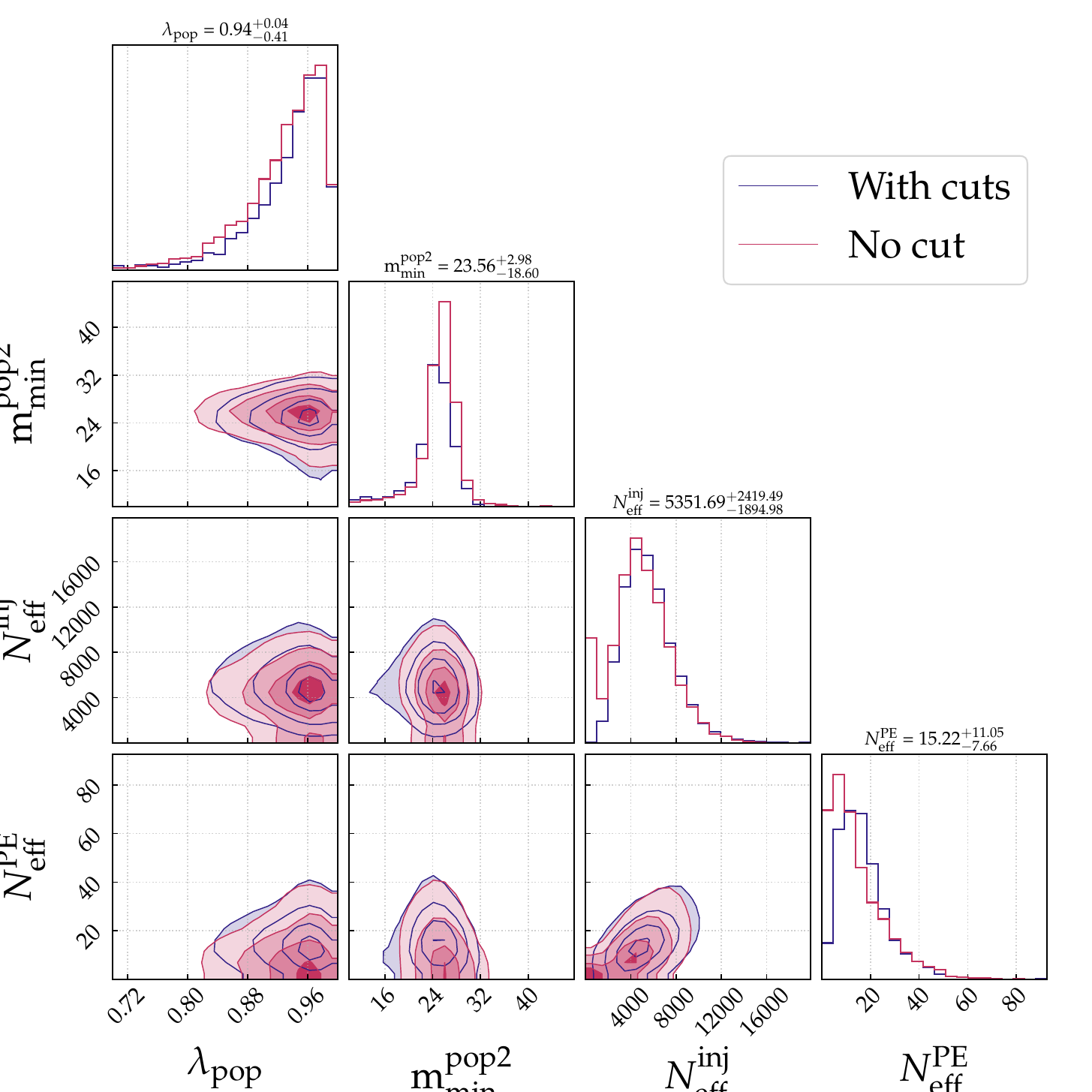}
    \caption{\textbf{Corner plot \textsc{Mixture model}:} Corner plot of the population parameters ($m_{\rm t}$,$\delta_{\rm m_{t}}$) and the stability estimator ($N_{\rm eff}^{\rm inj}$,$N_{\rm eff}^{\rm PE}$), obtained with the \textsc{Mixture Vanilla} evolving model on the 59 GW events with $\rm IFAR \geq 1 yr$. The purple contours and histograms were estimated while putting the minimum value of $N_{\rm eff}^{\rm inj}=200$ and $N_{\rm eff}^{\rm PE}=10$. The pink contours and histograms were estimated while placing not cuts on these estimators. }
    \label{fig:corner mixture}
\end{figure*}

To test the numerical stability of our results, we conducted two supplementary analyses without applying cuts to our stability estimators: the effective number of samples for the GW events ($\rm N_{\rm eff}^{\rm PE}$) and the effective number of injections ($\rm N_{\rm eff}^{\rm inj}$). In the main analysis, these parameters were set to $\rm N_{\rm eff}^{\rm PE}=10$ and $\rm N_{\rm eff}^{\rm inj}=4 \rm N_{\rm GW}$. Figures \ref{fig:corner BtB} and \ref{fig:corner mixture} show the estimated distributions of key population parameters alongside the two stability estimators. We compared the estimated posteriors using the same 59 GW events with the \textsc{Beta to Beta} and \textsc{Mixture Vanilla} models, respectively. We do not observe any significant correlation between the population parameters and the stability estimators, indicating that the spin evolution part of the parameter space is not influenced by $\rm N_{\rm eff}^{\rm inj}$ or $\rm N_{\rm eff}^{\rm PE}$. Furthermore, removing these cuts did not cause the hierarchical likelihood to shift to a different part of the parameter space, as the posterior distributions remained very similar with and without the stability cuts. We conclude that the results obtained are stable with respect to the numerical stability of the hierarchical Bayesian inference.

\newpage
\clearpage

\onecolumngrid

\section{Population models and prior ranges}

\subsection{Vanilla analysis}
\begin{table*}[h!]
    \centering
    \begin{tabular}{ c p{10cm} p{2mm} p{3cm} }
    \hline\hline \\
        {\textbf{Parameter}} & \textbf{Description} &  & \textbf{Prior} \\ \\\hline\hline \\
        & {\textbf{Masses: PowerLaw plus Peak}}  & \\ \\
        $\alpha$ & Spectral index for the PL of the primary mass distribution. &  & $\mathcal{U}$($0$, $8$)\\
        $\beta$ & Spectral index for the PL of the mass ratio distribution. &  & $\mathcal{U}$($-1$, $10$)\\
        $\rm m_{min}$ & Minimum mass of the primary mass distribution $[\Msol]$. &  & $\mathcal{U}$($1 \Msol$, $8 \Msol$)\\
        $\rm m_{max}$ &  Maximum mass of the primary mass distribution $[\Msol]$. &  & $\mathcal{U}$($70 \Msol$, $130 \Msol$)\\
        $\lambda_{\rm g}$ & Fraction of the model in the Gaussian component. &  & $\mathcal{U}$($0$, $1$) \\
        $\mu_{\rm g}$ & Mean of the Gaussian in the primary mass distribution $[\Msol]$.  &  & $\mathcal{U}$($20 \Msol$, $50 \Msol$) \\
        $\sigma_{\rm g}$ & Width of the Gaussian  in the primary mass distribution $[\Msol]$.  &  & $\mathcal{U}$($1 \Msol$, $10 \Msol$)\\
        $\delta_{m}$ & Range of mass tapering at the lower end of the mass distribution $[\Msol]$.  &  & $\mathcal{U}$($1 \Msol$, $10 \Msol$)\\ \\
        \hline \\
        &{\textbf{Spins: Default model}} &\\ \\
        $\alpha^{\rm pop1}_{\chi}$ & First parameter of the Beta distribution for the spin magnitude. &  & $\mathcal{U}$($1$, $10$) \\
        $\beta^{\rm pop1}_{\chi}$ & Second parameter of the Beta distribution for the spin magnitude. &  & $\mathcal{U}$($1$, $10$)\\
        $\sigma^{\rm pop1}_{t}$ & Standard deviation of the truncated gaussian for the cosine tilt angle distribution. &  & $\mathcal{U}$($0$, $5$)\\
        $\xi^{\rm pop1}$ & Mixing parameter for the cosine of the tilt angle distribution. &  & $\mathcal{U}$($0$, $1$)\\ \\
        \hline \\
        &{\textbf{Rate: MD}} &\\ \\
        $\gamma$ & Slope of the power law regime before the point $z_p$. &  & $\mathcal{U}$($0$, $8$) \\
        $k$ & Slope of the power law regime after the point $z_{\rm p}$. &  & $\mathcal{U}$($0$, $8$)\\
        $z_p$ & Redshift turning point between the power law regimes .&  & $\mathcal{U}$($0$, $8$)\\ 
        $\mathcal{R}_{0}$ & Local value of the CBC merger rate, at $z=0$. & & $\mathcal{U}$(0,100) \\ \\
        \hline
    \end{tabular}
    \caption{Prior ranges and population parameters for the non evolving Vanilla analysis, construct with a Powerlaw plus Peak for the masses, a Default spin model and a MD CBC merger rate.}
  \label{tab: Vanilla analysis}
\end{table*}

\subsection{Mass prior ranges}

\begin{table*}[h!]
    \centering
    \begin{tabular}{ c p{10cm} p{2mm} p{3cm} }
    \hline\hline \\
        {\textbf{Parameter}} & \textbf{Description} &  & \textbf{Prior} \\ \\\hline\hline \\
        & {\textbf{Pop1: PowerLaw plus Peak}}  & \\ \\
        $\alpha^{\rm pop1}$ & Spectral index for the PL of the primary mass distribution. &  & $\mathcal{U}$($0$, $8$)\\
        $\beta^{\rm pop1}$ & Spectral index for the PL of the mass ratio distribution. &  & $\mathcal{U}$($-1$, $8$)\\
        $\rm m^{\rm pop1}_{min}$ & Minimum mass of the primary mass distribution $[\Msol]$. &  & $\mathcal{U}$($1 \Msol$, $8 \Msol$)\\
        $\rm m^{\rm pop1}_{max}$ &  Maximum mass of the primary mass distribution $[\Msol]$. &  & $\mathcal{U}$($45 \Msol$, $60 \Msol$)\\
        $\lambda^{\rm pop1}_{\rm g}$ & Fraction of the model in the Gaussian component. &  & $\mathcal{U}$($0$, $1$) \\
        $\mu^{\rm pop1}_{\rm g}$ & Mean of the Gaussian in the primary mass distribution $[\Msol]$.  &  & $\mathcal{U}$($20 \Msol$, $50 \Msol$) \\
        $\sigma^{\rm pop1}_{\rm g}$ & Width of the Gaussian  in the primary mass distribution $[\Msol]$.  &  & $\mathcal{U}$($1 \Msol$, $10 \Msol$)\\
        $\delta^{\rm pop1}_{m}$ & Range of mass tapering at the lower end of the mass distribution $[\Msol]$.  &  & $\mathcal{U}$($1 \Msol$, $10 \Msol$)\\ \\
        \hline \\
        &{\textbf{Pop2: PowerLaw}} &\\ \\
        $\alpha^{\rm pop2}$ & Spectral index for the PL of the primary mass distribution. &  & $\mathcal{U}$($0$, $8$)\\
        $\beta^{\rm pop2}$ & Spectral index for the PL of the mass ratio distribution. &  & $\mathcal{U}$($-1$, $8$)\\
        $\rm m^{\rm pop2}_{min}$ & Minimum mass of the primary mass distribution $[\Msol]$. &  & $\mathcal{U}$($1\Msol$, $8 \Msol$)\\
        $\rm m^{\rm pop2}_{max}$ &  Maximum mass of the primary mass distribution $[\Msol]$. &  & $\mathcal{U}$($80 \Msol$, $130 \Msol$)\\
        $\delta^{\rm pop2}_{m}$ & Range of mass tapering at the lower end of the mass distribution $[\Msol]$.  &  & $\mathcal{U}$($1 \Msol$, $10 \Msol$)\\ \\
        \hline \\
        &{\textbf{Common parameters}} &\\ \\
        $\lambda_{\rm pop}$ & Fraction of the population $\rm pop1$ w.r.t to the overall population. & & $\mathcal{U}$(0,1)\\ \\
        \hline
    \end{tabular}
    \caption{The same mass model and prior ranges are used for the \textsc{Mixture} vanilla and paired models.}
  \label{tab: Mass Mixture vanilla model}
\end{table*}

\begin{table*}[h!]
    \centering
    \begin{tabular}{ c p{10cm} p{2mm} p{3cm} }
    \hline\hline \\
        {\textbf{Parameter}} & \textbf{Description} &  & \textbf{Prior} \\ \\\hline\hline \\
        & {\textbf{Pop1: PowerLaw plus Peak}}  & \\ \\
        $\alpha^{\rm pop1}$ & Spectral index for the PL of the primary mass distribution. &  & $\mathcal{U}$($0$, $8$)\\
        $\beta^{\rm pop1}$ & Spectral index for the PL of the mass ratio distribution. &  & $\mathcal{U}$($-1$, $8$)\\
        $\rm m^{\rm pop1}_{min}$ & Minimum mass of the primary mass distribution $[\Msol]$. &  & $\mathcal{U}$($1\Msol$, $8 \Msol$)\\
        $\rm m^{\rm pop1}_{max}$ &  Maximum mass of the primary mass distribution $[\Msol]$. &  & $\mathcal{U}$($80 \Msol$, $130 \Msol$)\\
        $\delta^{\rm pop1}_{m}$ & Range of mass tapering at the lower end of the mass distribution $[\Msol]$.  &  & $\mathcal{U}$($1 \Msol$, $10 \Msol$)\\ \\
        \hline \\
        &{\textbf{Pop2: PowerLaw}} &\\ \\
        $\mu^{\rm pop2}$ & Mean of the gaussian & & $\mathcal{U}$($20\Msol$,$50\Msol$)\\
        $\sigma^{\rm pop2}$ & Standard deviation of the gaussian & & $\mathcal{U}$($2\Msol$,$10 \Msol$)\\ \\
        \hline \\
        &{\textbf{Common parameters}} &\\ \\
        $\lambda_{\rm pop}$ & Fraction of the population $\rm pop1$ w.r.t to the overall population. & & $\mathcal{U}$(0,1)\\ \\
        \hline
    \end{tabular}
    \caption{\textsc{Mixture} peak model.}
  \label{tab: Mass Mixture peak model}
\end{table*}

\begin{table*}[h!]
    \centering
    \begin{tabular}{ c p{10cm} p{2mm} p{3cm} }
    \hline\hline \\
        {\textbf{Parameter}} & \textbf{Description} &  & \textbf{Prior} \\ \\\hline\hline \\
        & {\textbf{PowerLaw plus Peak}}  & \\ \\
        $\alpha$ & Spectral index for the PL of the primary mass distribution. &  & $\mathcal{U}$($0$, $8$)\\
        $\beta$ & Spectral index for the PL of the mass ratio distribution. &  & $\mathcal{U}$($-1$, $10$)\\
        $\rm m_{min}$ & Minimum mass of the primary mass distribution $[\Msol]$. &  & $\mathcal{U}$($1 \Msol$, $8 \Msol$)\\
        $\rm m_{max}$ &  Maximum mass of the primary mass distribution $[\Msol]$. &  & $\mathcal{U}$($70 \Msol$, $130 \Msol$)\\
        $\lambda_{\rm g}$ & Fraction of the model in the Gaussian component. &  & $\mathcal{U}$($0$, $1$) \\
        $\mu_{\rm g}$ & Mean of the Gaussian in the primary mass distribution $[\Msol]$.  &  & $\mathcal{U}$($20 \Msol$, $50 \Msol$) \\
        $\sigma_{\rm g}$ & Width of the Gaussian  in the primary mass distribution $[\Msol]$.  &  & $\mathcal{U}$($1 \Msol$, $10 \Msol$)\\
        $\delta_{m}$ & Range of mass tapering at the lower end of the mass distribution $[\Msol]$.  &  & $\mathcal{U}$($1 \Msol$, $10 \Msol$)\\ \\
        \hline
    \end{tabular}
    \caption{The same mass model and prior ranges are used for the three \textsc{Evolving} models.}
  \label{tab: Mass Evolving and shifting}
\end{table*}

\subsection{CBC merger rate prior ranges}

\begin{table*}[h!]
    \centering
    \begin{tabular}{ c p{10cm} p{2mm} p{3cm} }
    \hline\hline \\
        {\textbf{Parameter}} & \textbf{Description} &  & \textbf{Prior} \\ \\\hline\hline \\
        & {\textbf{Pop1: Madau\&Dickinson rate}}  & \\ \\
        $\gamma^{\rm pop1}$ & Slope of the power law regime before the point $z_p$. &  & $\mathcal{U}$($0$, $8$) \\
        $k^{\rm pop1}$ & Slope of the power law regime after the point $z_{\rm p}$. &  & $\mathcal{U}$($0$, $8$)\\
        $z_p^{\rm pop1}$ & Redshift turning point between the power law regimes .&  & $\mathcal{U}$($0$, $8$)\\ \\
        \hline \\
        &{\textbf{Pop2: Madau\&Dickinson rate}} &\\ \\
        $\gamma^{\rm pop2}$ & Slope of the power law regime before the point $z_p$. &  & $\mathcal{U}$($0$, $8$) \\
        $k^{\rm pop2}$ & Slope of the power law regime after the point $z_{\rm p}$. &  & $\mathcal{U}$($0$, $8$)\\
        $z_p^{\rm pop2}$ & Redshift turning point between the power law regimes. &  & $\mathcal{U}$($0$, $8$)\\ \\
        \hline \\
        &{\textbf{Common parameters}} &\\ \\
        $\mathcal{R}_{0}$ & Local value of the CBC merger rate, at $z=0$. & & $\mathcal{U}$(0,100)\\ \\
        \hline
    \end{tabular}
    \caption{The same CBC merger rate models and prior ranges are used for all the three flavours of the \textsc{Mixture} models.}
  \label{tab: Rate Mixture vanilla model}
\end{table*}

\begin{table*}[h!]
    \centering
    \begin{tabular}{ c p{10cm} p{2mm} p{3cm} }
    \hline\hline \\
        {\textbf{Parameter}} & \textbf{Description} &  & \textbf{Prior} \\ \\\hline\hline \\
        & {\textbf{Madau\&Dickinson rate}}  & \\ \\
        $\gamma$ & Slope of the power law regime before the point $z_p$. &  & $\mathcal{U}$($0$, $8$) \\
        $k$ & Slope of the power law regime after the point $z_{\rm p}$. &  & $\mathcal{U}$($0$, $8$)\\
        $z_p$ & Redshift turning point between the power law regimes .&  & $\mathcal{U}$($0$, $8$)\\ 
        $\mathcal{R}_{0}$ & Local value of the CBC merger rate, at $z=0$. & & $\mathcal{U}$(0,100) \\ \\
        \hline
    \end{tabular}
    \caption{The same CBC merger rate model and prior ranges are used for all the three flavours of \textsc{Evolving} models.}
  \label{tab: Rate Evolving models}
\end{table*}

\subsection{Spin prior ranges}

\begin{table*}[h!]
    \centering
    \begin{tabular}{ c p{10cm} p{2mm} p{3cm} }
    \hline\hline \\
        {\textbf{Parameter}} & \textbf{Description} &  & \textbf{Prior} \\ \\\hline\hline \\
        & {\textbf{Pop1: Default spin}}  & \\ \\
        $\alpha^{\rm pop1}_{\chi}$ & First parameter of the Beta distribution for the spin magnitude. &  & $\mathcal{U}$($1$, $10$) \\
        $\beta^{\rm pop1}_{\chi}$ & Second parameter of the Beta distribution for the spin magnitude. &  & $\mathcal{U}$($1$, $10$)\\
        $\sigma^{\rm pop1}_{t}$ & Standard deviation of the truncated gaussian for the cosine tilt angle distribution. &  & $\mathcal{U}$($0$, $5$)\\
        $\xi^{\rm pop1}$ & Mixing parameter for the cosine of the tilt angle distribution. &  & $\mathcal{U}$($0$, $1$)\\ \\
        \hline \\
        &{\textbf{Pop2: Default spin}} &\\ \\
        $\alpha^{\rm pop2}_{\chi}$ & First parameter of the Beta distribution for the spin magnitude. &  & $\mathcal{U}$($1$, $10$) \\
        $\beta^{\rm pop2}_{\chi}$ & Second parameter of the Beta distribution for the spin magnitude. &  & $\mathcal{U}$($1$, $10$)\\
        $\sigma^{\rm pop2}_{t}$ & Standard deviation of the truncated gaussian for the cosine tilt angle distribution. &  & $\mathcal{U}$($0$, $5$)\\
        $\xi^{\rm pop2}$ & Mixing parameter for the cosine of the tilt angle distribution. &  & $\mathcal{U}$($0$, $1$)\\ \\
        \hline
    \end{tabular}
    \caption{The same spin models and prior ranges are used for all the three flavours of the \textsc{Mixture} models.}
  \label{tab: Spin Mixture vanilla model}
\end{table*}

\begin{table*}[h!]
    \centering
    \begin{tabular}{ c p{10cm} p{2mm} p{3cm} }
    \hline\hline \\
        {\textbf{Parameter}} & \textbf{Description} &  & \textbf{Prior} \\ \\\hline\hline \\
        & {\textbf{Evolving Gaussian}}  & \\ \\
        $\mu_{\chi}$ & Zero order parameter expansion of the mean of the gaussian for the spin magnitude&  & $\mathcal{U}$($0$, $1$) \\
        $\sigma_{\chi}$ & Zero order parameter expansion of the standard deviation of the gaussian for the spin magnitude&  & $\mathcal{U}$($10^{-3}$, $2$)\\
        $\dot{\mu}_{\chi}$ & First order parameter expansion of the mean of the  gaussian for the spin magnitude. &  & $\mathcal{U}$($0$, $0.1$)\\
        $\dot{\sigma}_{\chi}$ & First order parameter expansion of the standard deviation of the  gaussian for the spin magnitude. &  & $\mathcal{U}$($-0.1$, $0.1$)\\ 
        $\sigma_{t}$ & Standard deviation of the truncated gaussian for the cosine tilt angle distribution. &  & $\mathcal{U}$($0$, $5$)\\
        $\xi$ & Mixing parameter for the cosine of the tilt angle distribution. &  & $\mathcal{U}$($0$, $1$)\\ \\
        \hline
    \end{tabular}
    \caption{Spin parameters and prior ranges for the \textsc{Evolving} model}
  \label{tab: Spin Evolving model}
\end{table*}

\begin{table*}[h!]
    \centering
    \begin{tabular}{ c p{10cm} p{2mm} p{3cm} }
    \hline\hline \\
        {\textbf{Parameter}} & \textbf{Description} &  & \textbf{Prior} \\ \\\hline\hline \\
        & {\textbf{Shifting Beta to Gaussian}}  & \\ \\
        $\alpha_{\chi}$ & First parameter of the Beta distribution for the spin magnitude. &  & $\mathcal{U}$($1$, $10$) \\
        $\beta_{\chi}$ & Second parameter of the Beta distribution for the spin magnitude. &  & $\mathcal{U}$(1,10)\\
        $\mu_{\chi}$ & Mean of the gaussian for the spin magnitude. &  & $\mathcal{U}$($0$, $1$) \\
        $\sigma_{\chi}$ & Standard deviation of the gaussian for the spin magnitude. &  & $\mathcal{U}$($10^{-3}$, $2$)\\
        $\sigma_{t}$ & Standard deviation of the truncated gaussian for the cosine tilt angle distribution. &  & $\mathcal{U}$($0$, $5$)\\
        $\xi$ & Mixing parameter for the cosine of the tilt angle distribution. &  & $\mathcal{U}$($0$, $1$)\\ 
        $\rm m_{t}$ & Critical mass at which the window function is equal to $0.5$ (transition point). &  & $\mathcal{U}$($10 M_{\odot}$, $100 M_{\odot}$)\\
        $\delta_{m_{t}}$ & Steepness of the window function. &  & $\mathcal{U}$($1$, $20$)\\ 
        $\rm f_{\rm mix}$ & Starting value of the window function. &  & $\mathcal{U}$($0$, $1$)\\ \\
        \hline
    \end{tabular}
    \caption{Spin parameters and prior ranges for the \textsc{Evolving} Beta to Gaussian}
  \label{tab: Spin Beta to Gaussian model}
\end{table*}

\begin{table*}[h!]
    \centering
    \begin{tabular}{ c p{10cm} p{2mm} p{3cm} }
    \hline\hline \\
        {\textbf{Parameter}} & \textbf{Description} &  & \textbf{Prior} \\ \\\hline\hline \\
        & {\textbf{Shifting Beta to Gaussian}}  & \\ \\
        $\alpha^{\rm low}_{\chi}$ & First parameter of the Beta distribution for the spin magnitude. &  & $\mathcal{U}$($1$, $10$) \\
        $\beta^{\rm low}_{\chi}$ & Second parameter of the Beta distribution for the spin magnitude. &  & $\mathcal{U}$(1,10)\\
        $\alpha^{\rm high}_{\chi}$ & First parameter of the Beta distribution for the spin magnitude. &  & $\mathcal{U}$($1$, $10$) \\
        $\beta^{\rm high}_{\chi}$ & Second parameter of the Beta distribution for the spin magnitude. &  & $\mathcal{U}$(1,10)\\
        $\sigma_{t}$ & Standard deviation of the truncated gaussian for the cosine tilt angle distribution. &  & $\mathcal{U}$($0$, $5$)\\
        $\xi$ & Mixing parameter for the cosine of the tilt angle distribution. &  & $\mathcal{U}$($0$, $1$)\\
        $\rm m_{t}$ & Critical mass at which the window function is equal to $0.5$ (transition point). &  & $\mathcal{U}$($10 M_{\odot}$, $100 M_{\odot}$)\\
        $\delta_{m_{t}}$ & Steepness of the window function. &  & $\mathcal{U}$($1$, $20$)\\ 
        $\rm f_{\rm mix}$ & Starting value of the window function. &  & $\mathcal{U}$($0$, $1$)\\ \\
        \hline
    \end{tabular}
    \caption{Spin parameters and prior ranges for the \textsc{Evolving} Beta to Beta}
  \label{tab: Spin Beta to Beta model}
\end{table*}

\end{document}